\documentclass{emulateapj}
\usepackage{natbib,times}
\usepackage{lscape}

\slugcomment{accepted for publication in ApJS on Feb. 29th 2012.}

\shorttitle{}
\shortauthors{}
\begin{document}

\title{A catalog of 132,684 clusters of galaxies identified from Sloan Digital Sky Survey III}

\author{Z. L. Wen\altaffilmark{1}, 
        J. L. Han\altaffilmark{1}
        and 
        F. S. Liu\altaffilmark{2}}

\altaffiltext{1}{National Astronomical Observatories, Chinese Academy of Sciences, 
                 20A Datun Road, Chaoyang District, Beijing 100012, China; 
                 zhonglue@nao.cas.cn.} 
\altaffiltext{2}{College of Physics Science and Technology, 
                 Shenyang Normal University, Shenyang 110034, China.}


\begin{abstract}

Using the photometric redshifts of galaxies from the Sloan Digital Sky
Survey III (SDSS-III), we identify 132,684 clusters in the redshift
range of $0.05\le z <0.8$. Monte Carlo simulations show that the false
detection rate is less than 6\% for the whole sample. The completeness
is more than 95\% for clusters with a mass of
$M_{200}>1.0\times10^{14}~M_{\odot}$ in the redshift range of $0.05\le
z< 0.42$, while clusters of $z>0.42$ are less complete and have a
biased smaller richness than the real one due to incompleteness of
member galaxies. We compare our sample with other cluster samples, and
find that more than 90\% of previously known rich clusters of $0.05\le
z <0.42$ are matched with clusters in our sample. Richer clusters tend
to have more luminous brightest cluster galaxies (BCGs). Correlating
with X-ray and the Planck data, we show that the cluster
richness is closely related to the X-ray luminosity, temperature, and
Sunyaev--Zel'dovich measurements.
Comparison of the BCGs with the SDSS luminous red galaxy (LRG)
sample shows that 25\% of LRGs are BCGs of our clusters and 36\% of
LRGs are cluster member galaxies. In our cluster sample, 63\% of 
BCGs of $r_{\rm petro}<19.5$ satisfy the SDSS LRG selection criteria.

\end{abstract}

\keywords{galaxies: clusters: general -- galaxies: distances and redshifts}

\section{Introduction}

Clusters of galaxies are usually located at the knots of the
filamentary structures in the universe. They are the most massive
bound systems to trace the large-scale structure
\citep{bah88,phg92,cye+96,bfc97}. Statistical studies of clusters
provide very powerful constraint on the cosmological parameters
\citep[see a review in][]{aem11} by using, e.g., cluster mass function
\citep{rb02,sel02,dah06,pd07,rdn07,whl10} and gas fraction in massive
clusters \citep{ars+08}.
Clusters are also important laboratories to investigate the evolution
of galaxies in dense environment \citep{dre80,bo78,bo84,gyf+03} and
act as natural telescope to study lensed high-redshift faint
background galaxies \citep{bki+99,sib+02,mkm+03,sek+04}. As a large
number of clusters have been detected \citep{kma+07b,whl09}, the
distributions of galaxy clusters are used to detect the baryon
acoustic oscillation of the universe \citep{esf09,hut10,hhw+12}.
Correlation of background objects with a large sample of clusters
shows the effects of weak lensing and spectra line absorption
\citep{mos+05, lbl+08}.

Galaxy clusters have been found from single-band optical imaging data
\citep[e.g.,][]{abe58,plg+96}, multicolor photometric data
\citep[e.g.,][]{gy05,kma+07b} and spectroscopic redshift surveys
\citep[e.g.,][]{hg82,ymv+05}. In addition, clusters have also been
found from X-ray surveys \citep[e.g.,][]{bvh+00,bsg+04}. At millimeter
wavelength, the Sunyaev--Zel'dovich (SZ) effect has recently used to
find clusters, which is insensitive to cluster redshift
\citep{cjg+00}. Hundreds of SZ clusters have been identified
\citep{maa+11,planck11}.

The Sloan Digital Sky Survey \citep[SDSS;][]{yaa+00} offers an
opportunity to produce the largest and most complete cluster
sample. It provides photometry in five broad bands ($u$, $g$, $r$,
$i$, and $z$) covering 14,000 deg$^2$ and the follow-up spectroscopic
observations. The photometric data reach a limit of $r=22.2$
\citep{slb+02} with the star--galaxy separation reliable to a limit of
$r=21.5$ \citep{lgi+01}. The spectroscopic survey observes galaxies
with an extinction-corrected Petrosian magnitude of $r<17.77$ for the
main galaxy sample \citep{swl+02} and $r<19.5$ for the luminous red
galaxy (LRG) sample \citep{eag+01}.
Galaxy clusters or groups have been found by using the SDSS
spectroscopic data \citep[e.g.,][]{mz05,bfw+06} and the photometric
data \citep[e.g.,][]{gsn+02,kma+07b,whl09,hmk+10,spd+11}.

For large samples of galaxy clusters, the determination of optical
richness still has a large uncertainty because of difficulties in the
discrimination of cluster member galaxies from background galaxies
using photometric data. This causes a large scatter when the optical
richness is used to represent the cluster mass for the constraint of
cosmological parameters \citep{rmb+08,rrk+09,whl10}.

In this paper, we improve the method of \citet{whl09} using
photometric redshifts to identify a large sample of galaxy clusters up
to $z\sim0.8$. In Section 2, we identify clusters using a new cluster
detection algorithm and determine a cluster richness that is closely
related to cluster mass. In Section 3, we compare our sample with the
previous cluster samples from the SDSS, and correlate the cluster
richness with X-ray and SZ measurements. In Section 4, we study the
brightest cluster galaxies (BCGs) and cross-identify them with the
SDSS LRGs. A summary is presented in Section 5.

Throughout this paper, we assume a $\Lambda$CDM cosmology, taking
$H_0=$100 $h$ ${\rm km~s}^{-1}$ ${\rm Mpc}^{-1}$, with $h=0.72$, 
$\Omega_m=0.3$ and $\Omega_{\Lambda}=0.7$.

\section{Clusters identified from SDSS-III}

\citet{whl09} identified 39,668 galaxy clusters from the SDSS DR6 by
the discrimination of member galaxies of clusters using photometric
redshifts (hereafter photo-$z$s) of galaxies. \citet{wh11} improved
the method and successfully identified the high-redshift clusters from the
deep fields of the Canada--France--Hawaii Telescope Wide (CFHT) survey,
the CHFT Deep survey, the Cosmic Evolution Survey, and the {\it Spitzer}
Wide-area InfraRed Extragalactic survey. Here, we follow and improve
the algorithm to identify clusters from SDSS-III \citep[SDSS Data
  Release 8;] []{dr8+11}. We first use the SDSS data to determine
scaling relations for cluster mass and radius which will be used in
the cluster detection.

\subsection{SDSS Data}

The galaxy data of 14,000 deg$^2$ are downloaded from the SDSS-III
database. The photo-$z$s, the $K$-corrections, and the absolute
magnitudes are obtained from the table {\it Photoz} in which
photo-$z$s are estimated based on the method of \citet{cdt+07}. We
remove objects with deblending problems and saturated objects using
the flags\footnote{(flags \& 0$\times$20) = 0 and (flags \&
  0$\times$80000) = 0 and ((flags \& 0$\times$400000000000) = 0 or
  psfmagerr$_r <= 0.20$) and ((flags \& 0$\times$40000) = 0).}.

\begin{figure}
\epsscale{1.}
\plotone{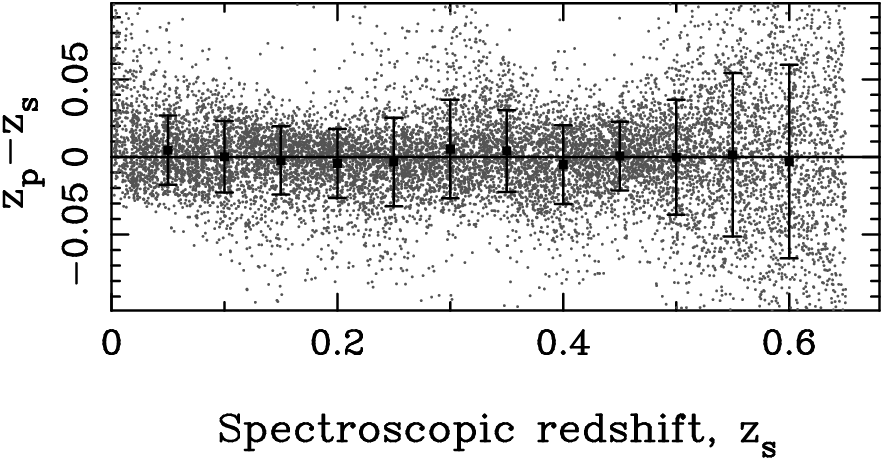}
\caption{Difference between spectroscopic redshift $z_{\rm s}$ and 
  photometric redshift $z_{\rm p}$ for galaxies of the SDSS-III.
\label{photozg}}
\end{figure}

As shown in Figure~\ref{photozg}, the uncertainties of photo-$z$s are
$\sim$0.025--0.030 in the redshift range $z<0.45$. They become larger
at higher redshifts. In the following analysis, we assume that the
uncertainty of photo-$z$, $\sigma_z$, increases with redshift in the
form of $\sigma_z=\sigma_0(1+z)$ for all galaxies. Here, we discard
those galaxies with a large photo-$z$ error $z{\rm Err}>0.08(1+z)$,
i.e., about $3\,\sigma_z$, which suffer bad photometry or
contamination of stars. This procedure removes 20\% of objects, most
of which are faint objects ($r>21$) with large photometric errors.

BCGs are the most luminous members of clusters. Generally, the BCGs are
elliptical galaxies and have smaller photo-$z$ error than other member
galaxies. Proper selection of BCGs can be helpful to identify clusters
and estimate the cluster parameters. To get right BCGs, we select
those galaxies as BCG candidates which have a photo-$z$ error $z{\rm
  Err}\le0.055(1+z)$ and a galaxy ellipticity in the $r$ band less
than 0.7.

\subsection{Scaling relations for cluster mass and radius}
\label{scaling}

\begin{figure}
\epsscale{1.}
\plotone{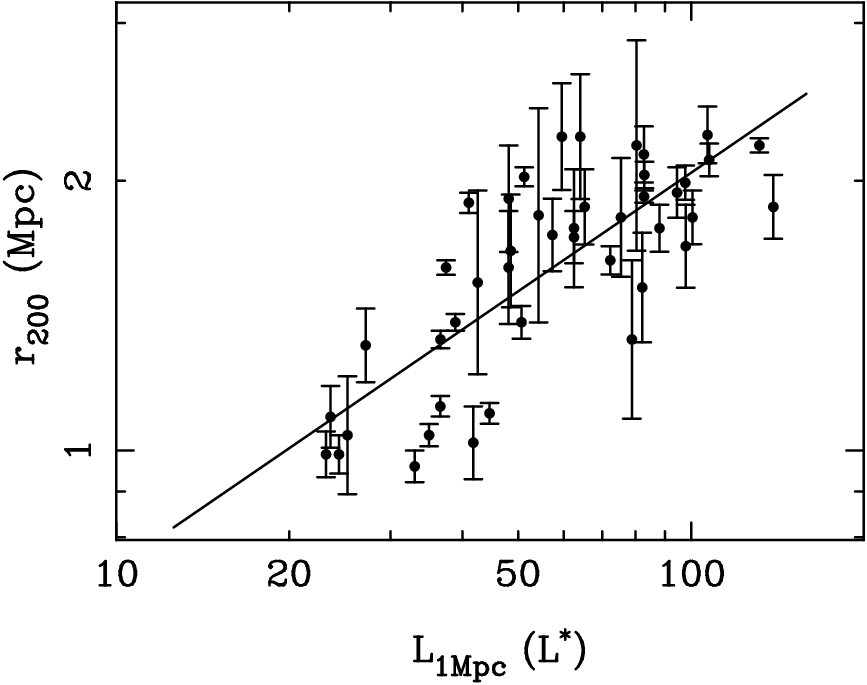}\\[2mm]
\plotone{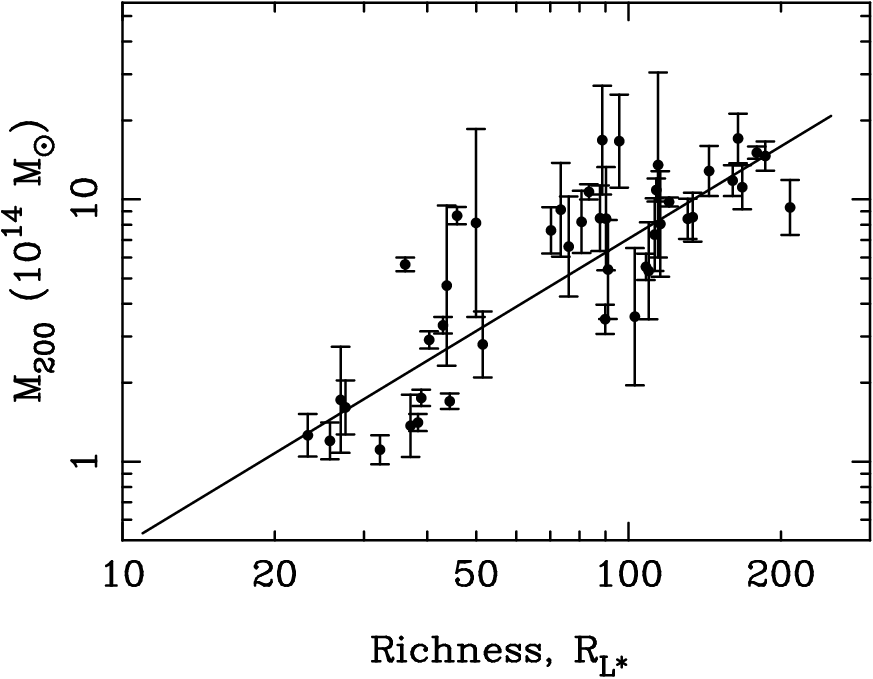}
\caption{Upper: scaling relation between cluster radius 
$r_{200}$ and total luminosity within a radius of 1 Mpc in units 
of $L^{\ast}$. Lower: scaling relation between cluster 
mass $M_{200}$ and richness $R_{L\ast}$. The solid lines show the 
best-fit relations. Data are derived for clusters compiled by \citet{whl10}.
\label{r200m200}}
\end{figure}

The cluster mass and radius are two fundamental parameters for
clusters. The widely used are $r_{200}$, the radius within which the
mean density of a cluster is 200 times of the critical density of the
universe, and $M_{200}$, the cluster mass within $r_{200}$.

We use the SDSS data of known clusters to get the scaling relations
for $r_{200}$ and $M_{200}$. Here, we take the clusters whose masses
or radii have been estimated by X-ray or weak-lensing methods as
compiled by \citet{whl10}. For each cluster, we calculate the total
luminosities of cluster member candidates in the SDSS $r$-band within
a radius of 1 Mpc from its BCG by summing luminosities of member
galaxy candidates brighter than $M^e_r(z)\le-20.5$ but fainter than
the BCG within a photo-$z$ gap of $z\pm 0.04(1+z)$, with a local
background subtraction. Here, $M^e_r$ is the evolution-corrected
absolute magnitude in the $r$ band, $M^e_r(z)=M_r(z)+Qz$, where we
adopt a passive evolution of $Q=1.62$ \citep{bhb+03}. To estimate
local background, we follow the method analogous to
\citet{pbb+04}. For each cluster, we divide the annuals between 2 and 4
Mpc from its BCG into 48 sections with equal area. Within the same
magnitude and photo-$z$ range, we calculate the total luminosity in
each sector and estimate the mean value and its root mean square. The
regions with luminosities larger than 3$\,\sigma$ are discarded and
the mean is recalculated to be the local background. We measure the
total luminosity of cluster member candidates in units of $L^{\ast}$,
here $L^{\ast}$ is the evolved characteristic luminosity of galaxies
in the $r$ band, defined as $L^{\ast}(z)=L^{\ast}(z=0)10^{0.4Qz}$
\citep{bhb+03}.

In Figure~\ref{r200m200}, we show the correlation between cluster
radius $r_{200}$ and total luminosity within a radius of 1 Mpc,
$L_{\rm 1Mpc}$. The best fit gives
\begin{equation}
\log r_{200}=(-0.57\pm0.09)+(0.44\pm0.05)\log L_{\rm 1Mpc},
\label{r200}
\end{equation}
where $r_{200}$ is in units of Mpc and $L_{\rm 1Mpc}$ is in units of
$L^{\ast}$. Similarly, we also get the total $r$-band luminosity
within the radius of $r_{200}$, $L_{200}$, in units of $L^{\ast}$,
with a background subtraction. In this paper, we define the cluster
richness as $R_{L\ast}=L_{200}/L^{\ast}$. We find that cluster mass
$M_{200}$ and cluster richness are closely correlated as (lower panel
of Figure~\ref{r200m200})
\begin{equation}
\log M_{200}=(-1.49\pm0.05)+(1.17\pm0.03)\log R_{L\ast},
\label{m200}
\end{equation}
where $M_{200}$ is in units of $10^{14}~M_{\odot}$. Using these
scaling relations, we can estimate the cluster radius and mass from
observables of $L_{\rm 1Mpc}$ and $L_{200}$ or $R_{L\ast}$.  We have
shown in our previous work that the measurements of the total
luminosity or richness are robust for adopting different widths of the
photo-$z$ gap \citep{whl09}. In the following, we extrapolate these
relations to lower richness for cluster identification. Based on the
estimated richness, we identify clusters with a richness threshold of
$R_{L\ast}\ge 12$.

\subsection{Cluster detection algorithm}
\label{algorithm}

\begin{deluxetable*}{rrrrrccrrl}[t]
\tablecolumns{10}
\tablewidth{0pc}
\tablecaption{Clusters of Galaxies Identified from the SDSS-III
\label{cat}}
\tablehead{ 
\colhead{Name}&\colhead{R.A.$_{\rm BCG}$}&\colhead{Decl.$_{\rm BCG}$} & \colhead{$z_{\rm p}$}  &
\colhead{$z_{\rm s,BCG}$} & \colhead{$r_{\rm BCG}$} & \colhead{$r_{200}$} & \colhead{$R_{L\ast}$}
& \colhead{$N_{200}$} & \colhead{Other Catalogs}\\
  &\colhead{(deg)}& \colhead{(deg)} & & & & \colhead{(Mpc)} & & & \\
\colhead{(1)} & \colhead{(2)} & \colhead{(3)} & \colhead{(4)} & 
\colhead{(5)} & \colhead{(6)} & \colhead{(7)} & \colhead{(8)} & 
\colhead{(9)} & \colhead{(10)} 
}
\startdata 
WHL J000000.6$+$321233& 0.00236&  32.20925& 0.1274& $-1.0000$& 14.92& 1.72& 70.63& 24&  Abell               \\             
WHL J000002.3$+$051718& 0.00957&   5.28827& 0.1696& $-1.0000$& 16.20& 0.94& 17.48&  9&                      \\             
WHL J000003.3$+$311354& 0.01377&  31.23175& 0.5428& $-1.0000$& 20.17& 0.87& 14.27&  8&                      \\             
WHL J000003.5$+$314708& 0.01475&  31.78564& 0.0932& $-1.0000$& 15.18& 0.94& 16.97&  9&                      \\             
WHL J000004.7$+$022826& 0.01945&   2.47386& 0.4179& $-1.0000$& 19.32& 0.95& 13.71& 10&                      \\             
WHL J000004.9$-$033248& 0.02024&$-3.54679$& 0.5968& $-1.0000$& 20.67& 1.00& 19.19& 11&                      \\             
WHL J000005.5$+$354610& 0.02303&  35.76957& 0.4762& $-1.0000$& 19.59& 0.91& 15.58&  9&                      \\             
WHL J000006.0$+$152548& 0.02482&  15.42990& 0.1656& $-1.0000$& 16.60& 1.13& 23.53& 19&  maxBCG,WHL09,GMBCG  \\             
WHL J000006.3$+$221220& 0.02643&  22.20558& 0.3985& $-1.0000$& 19.36& 0.84& 12.73& 11&                      \\             
WHL J000006.6$+$100648& 0.02755&  10.11333& 0.3676& $-1.0000$& 19.07& 0.93& 16.73& 13&                      \\             
WHL J000006.6$+$315235& 0.02762&  31.87626& 0.2134& $-1.0000$& 17.11& 1.18& 28.58& 15&                      \\             
WHL J000006.6$+$292129& 0.02765&  29.35813& 0.2489& $-1.0000$& 18.13& 0.92& 15.11& 16&                      \\             
WHL J000007.1$-$092910& 0.02957&$-9.48607$& 0.3332& $-1.0000$& 19.11& 0.80& 12.81& 13&  AMF                 \\             
WHL J000007.6$+$155003& 0.03177&  15.83424& 0.1436&   0.1528 & 15.99& 1.17& 34.11& 27&  Abell,maxBCG,WHL09,GMBCG,AMF \\    
WHL J000007.7$+$185245& 0.03208&  18.87909& 0.4347& $-1.0000$& 18.95& 1.03& 20.47& 12&                      \\
       
\enddata
\tablecomments{
Column 1: Cluster name with J2000 coordinates of cluster; 
Column 2: R.A. (J2000) of cluster and BCG; 
Column 3: Decl. (J2000) of cluster and BCG; 
Column 4: photometric redshift of cluster; 
Column 5: spectroscopic redshift of BCG, $-1.0000$ means not available; 
Column 6: $r$-band magnitude of BCG; 
Column 7: $r_{200}$ of cluster (Mpc);
Column 8: cluster richness;
Column 9: number of member galaxy candidates within $r_{200}$;
Column 10: other catalogs containing the cluster: 
Abell \citep{abe58,aco89}; maxBCG \citep{kma+07b}; WHL09 \citep{whl09}; 
GMBCG \citep{hmk+10}; AMF \citep{spd+11}.\\
(This table is available in its entirety in a machine-readable form in the online journal. 
A portion is shown here for guidance regarding its form and content.)
}
\end{deluxetable*}

In \citet{whl09}, we identified a cluster if more than eight member
galaxies of $M_r\le -21$ (without evolution correction and different
from $M^e_r(z)$ in this paper) are found within a radius of 0.5 Mpc
and a photo-$z$ gap of $z\pm0.04(1+z)$. The richness was estimated to
be the number of galaxies within a radius of 1 Mpc which may be
systematically smaller than the true radius for rich clusters but
larger than that for poor clusters. Note also that the radius of 0.5
Mpc for cluster identification is not the same as the radius of 1 Mpc
for richness estimate.

With the tight scaling relations shown in Section~\ref{scaling}, we
can improve the method of \citet{whl09} and \citet{wh11} to identify
clusters from the SDSS-III. Our new algorithm includes following
steps.

1. For each galaxy at a given photometric redshift, $z$, we count the
number of luminous member galaxies of $M^e_r(z)\le-20.5$ within a
radius of 0.5 Mpc and a photo-$z$ gap of $z\pm0.04(1+z)$,
$N_{\rm 0.5Mpc}$.

2. To get cluster candidates, we apply the friend-of-friend algorithm
\citep{hg82} to the luminous galaxies using a linking length of 0.5
Mpc in the transverse direction and a photo-$z$ difference of
$0.06(1+z)$. The linked galaxy with the maximum $N_{\rm 0.5Mpc}$ is
taken as the temporary center of a cluster candidate. If two or more
galaxies have the same maximum number, the brightest one is taken as
the temporary central cluster galaxy. Then, we obtain a temporary list
of cluster candidates.

3. For each cluster candidate at $z$, we assume that the galaxies of
$M^e_r(z)\le-20.5$ within a radius of 1 Mpc from the temporary central
galaxy and the photo-$z$ gap of $z\pm0.04(1+z)$ are member
galaxies. The cluster redshift $z_{\rm p}$ is then defined to be the
median value of the photometric redshifts of the recognized
``members''.

4. We recognize the BCG of a cluster candidate as the brightest galaxy
from the BCG candidate sample within a radius of 0.5 Mpc from the
temporary center and the photo-$z$ gap of $z\pm0.04(1+z)$. As a new
development in this paper, we use the BCG as the center of a cluster and
calculate $L_{\rm 1Mpc}$ from the estimated redshift. Subsequently, we
get $r_{200}$ using Equation~(\ref{r200}) and measure $R_{L\ast}$ from
$L_{200}$ within the estimated $r_{200}$ for each cluster
candidate. Note that the local background has been subtracted already
here.

5. We define a galaxy cluster if $R_{L\ast}\ge12$, which corresponds
to $M_{200}\sim0.6\times10^{14}~M_{\odot}$ using
Equation~(\ref{m200}). Note that a cluster candidate at low richness
may be contaminated by a small number of bright field galaxies with
their photo-$z$s seriously overestimated. To avoid such contamination,
we require the number of galaxies within a radius of $r_{200}$ and a
photo-$z$ gap of $z\pm0.04(1+z)$, $N_{200}\ge8$. This assures that the
cluster sample above the detection criteria has a high completeness
and a low false detection rate (see Sections~\ref{complete} and
\ref{false}).
After we get the cluster candidate list, we merge possible repeatedly
identified clusters which may survive in the previous procedure by
using the friend-of-friend algorithm. If two cluster candidates have
photo-$z$ difference less than $0.06(1+z)$ and a projection separation
less than $r_{200}$, we merge the cluster candidates as one cluster,
and the poorer one is then removed from the list.

\begin{figure}
\epsscale{1.}
\plotone{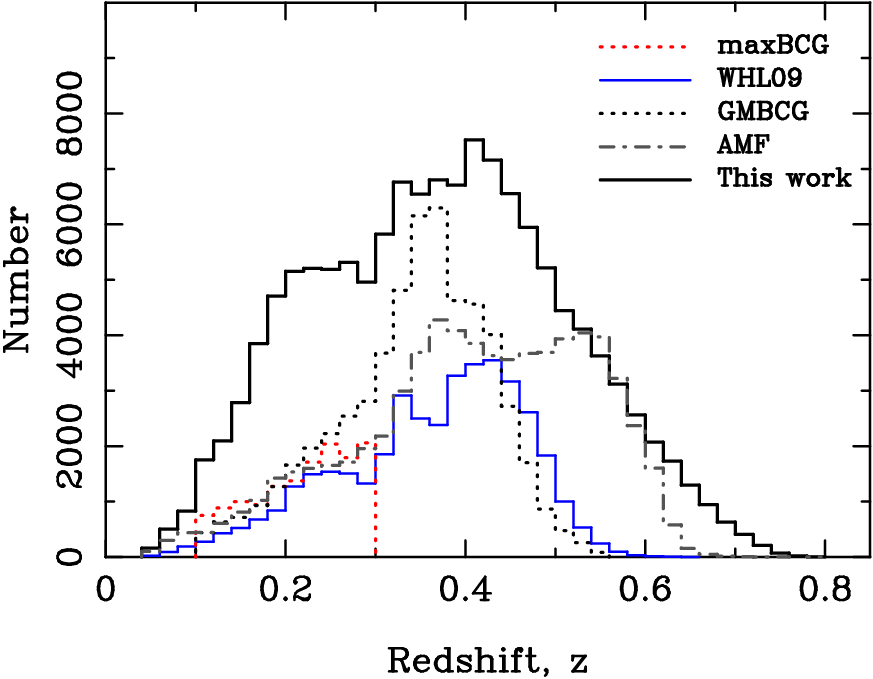}
\caption{Redshift distribution of the 132,684 identified clusters in
  the SDSS-III together with those of 13,823 maxBCG clusters within
  $0.1\le z\le 0.3$ \citep{kma+07b}, 39,668 WHL09 clusters within
  $0.05<z<0.6$ \citep{whl09}, 55,424 Gaussian Mixture
Brightest Cluster Galaxy (GMBCG) clusters within
  $0.1<z<0.55$ \citep{hmk+10} and 69,173 adaptive
matched-filter (AMF) clusters within $0.045\le
  z<0.78$ \citep{spd+11}.\\ (A color version of this figure is
  available in the online journal.)
\label{histcz}}
\end{figure}

Finally, we visually inspect the color images on the SDSS Web
site\footnote {http://skyserver.sdss3.org/dr8/en/tools/chart/list.asp}
for all cluster candidates. About 5000 (3.6\%) cluster candidates are
obvious contaminations from bad photometries and are removed. By this
procedure, we discovered 68 lensing systems from the color images
\citep{whj11}.

After above procedures, we find 132,684 clusters in the redshift of
$0.05\le z<0.8$ from the SDSS-III data. All clusters are listed in
Table~\ref{cat}. Figure~\ref{histcz} show the redshift distribution of
the identified clusters in the SDSS-III which is compared with those
of the maxBCG \citep{kma+07b}, WHL09 \citep{whl09}, Gaussian Mixture
Brightest Cluster Galaxy \citep[GMBCG;][]{hmk+10} and adaptive
matched-filter \citep[AMF;][]{spd+11} samples. Our new sample has a
peak at $z\sim 0.42$ and the number of our clusters is nearly two
times as the previously largest AMF sample by \citet{spd+11}. The
clusters of $z<0.42$ seem to be a complete sample. Because of the
magnitude limit of the SDSS photometric data, the member galaxies of
$z>0.42$ are incomplete for the threshold of $M^e_r(z)=-20.5$, so that
clusters of $z>0.42$ have a biased smaller richness than the real one
because of missing members. In the following, we only consider
clusters of $z<0.42$ when we compare our richness with other richness
estimates and X-ray measurements.

\newpage

\subsection{Redshift uncertainty}

\begin{figure}
\epsscale{1.1}
\plotone{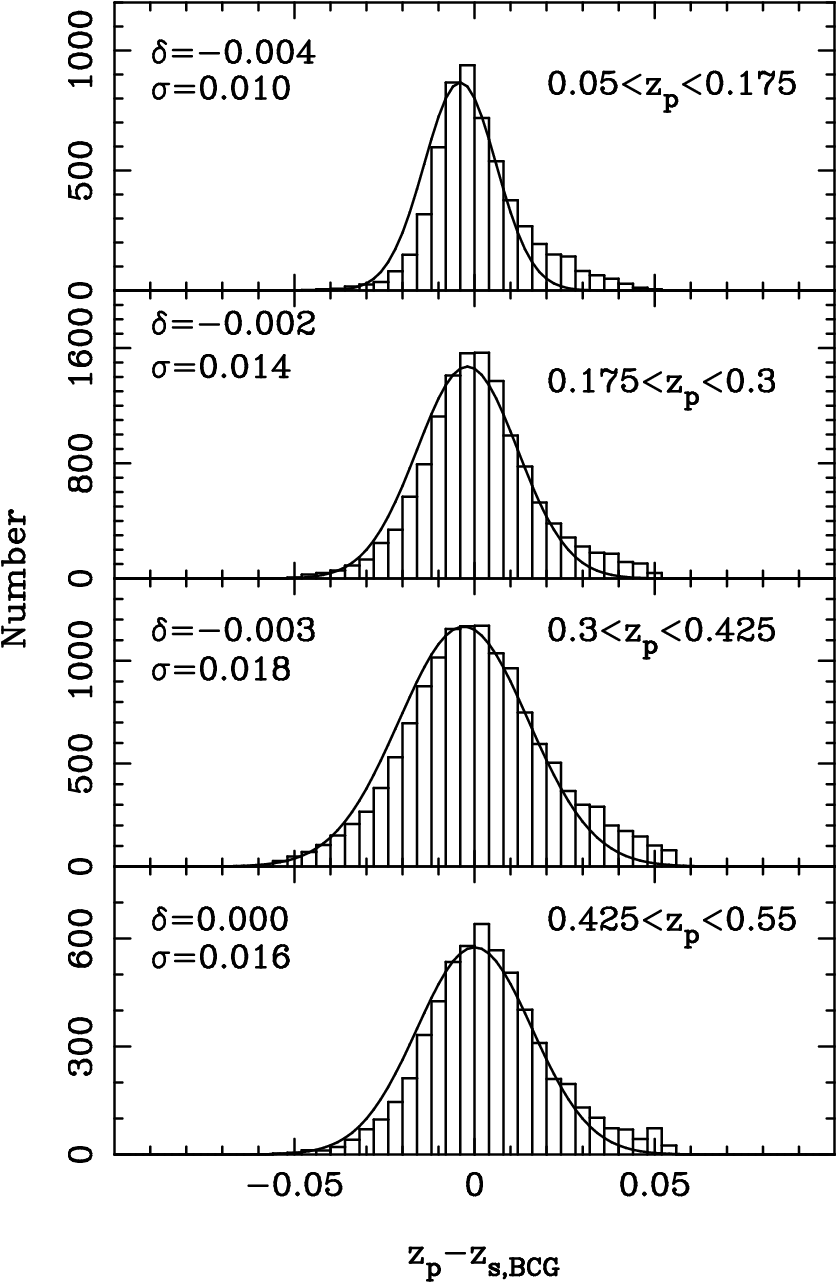}
\caption{Distribution of the difference between photometric and
  spectroscopic redshifts in the four redshift ranges. The solid line
  is the best fit with a Gaussian function. The parameters, i.e., the
  offset $\delta$ and the standard deviation $\sigma$, of the Gaussian
  function are marked on the left of each panel.
\label{phzc}}
\end{figure}

Using the SDSS spectroscopic data, we can get the spectroscopic
redshift of identified clusters and verify the accuracy of their
photometric redshifts. The spectroscopic redshift of a cluster is
taken to be that of its BCG. We find that the BCGs of 38,116 clusters
have spectroscopic redshifts.

Figure~\ref{phzc} shows the distribution of the difference between
photometric and spectroscopic redshifts, $z_{\rm p}-z_{\rm s,BCG}$. The
standard deviations roughly indicate the accuracy of redshift estimate
for clusters. In each panel, we fit the distribution of $z_{\rm
  p}-z_{\rm s,BCG}$ with a Gaussian function. The systematic offset
$|\delta|$ of the fitting is less than $0.004$, and the standard
deviation $\sigma$ is less than 0.018.

\subsection{Completeness of cluster detection}
\label{complete}

\begin{figure}
\epsscale{1.}
\plotone{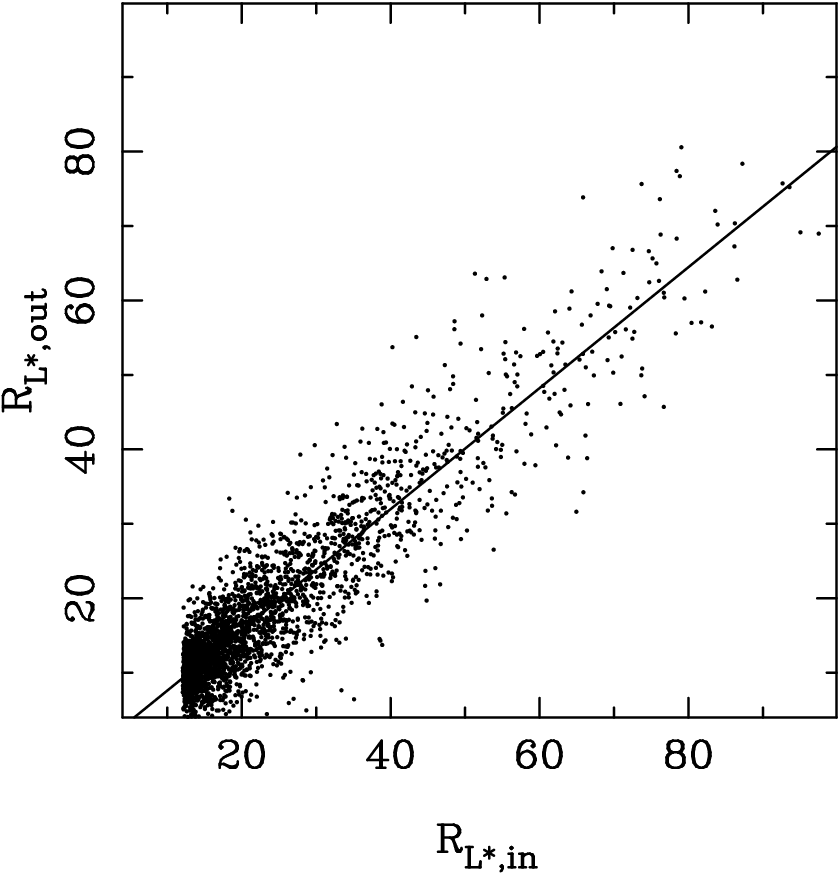}
\caption{Correlation between input richness and output richness
for the mock clusters. The solid line shows the best-fit relation.
\label{outrich}}
\end{figure}

\begin{figure}
\epsscale{1.}
\plotone{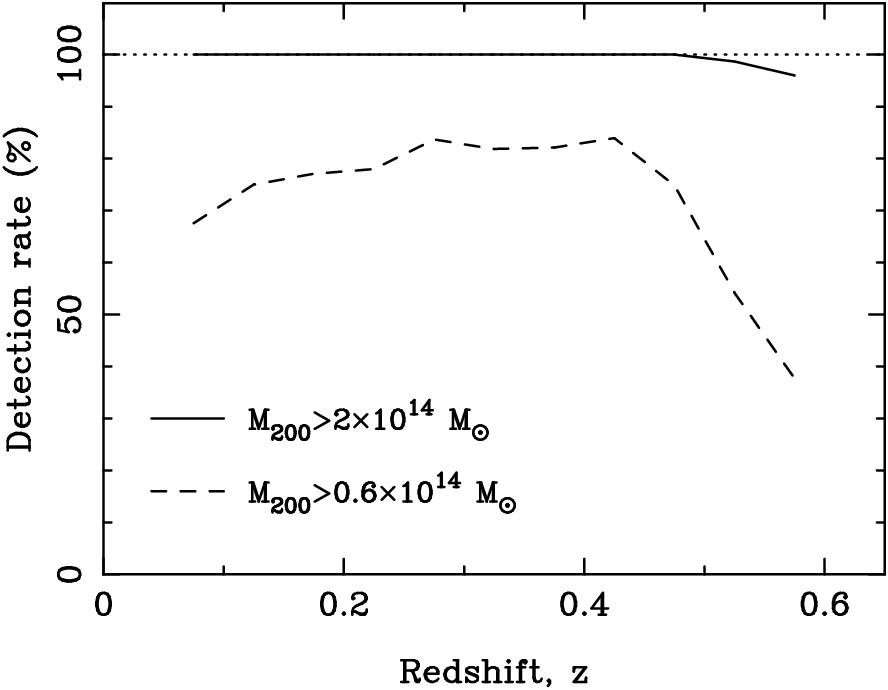}\\[2mm]
\plotone{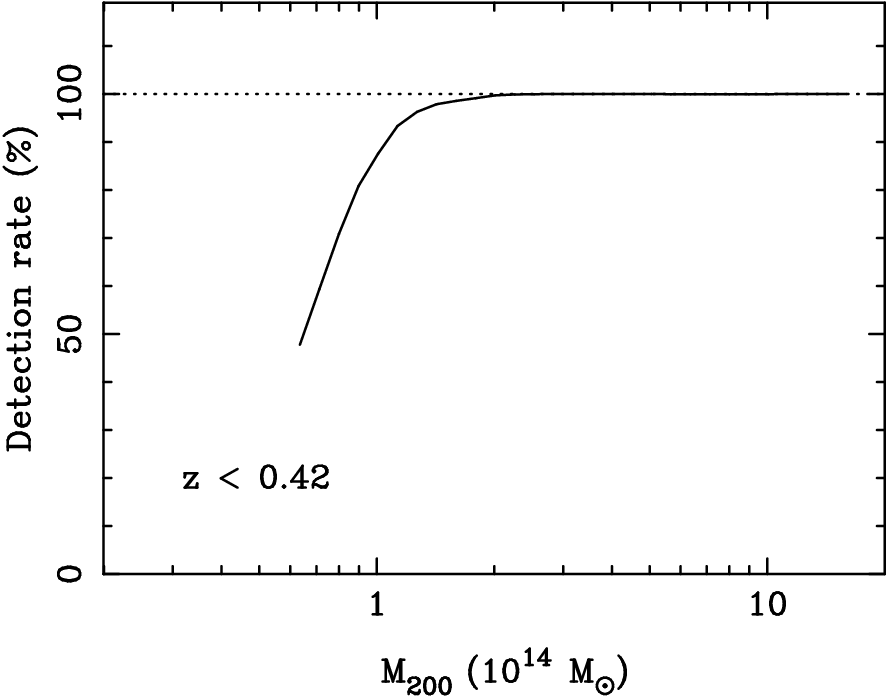}
\caption{Detection rate of the mock clusters as a 
function of redshift (the upper panel) and cluster mass, 
$M_{200}$ (the lower panel).
\label{detrate}}
\end{figure}

We use Monte Carlo simulation to test the completeness of our sample.
Mock clusters are simulated with assumptions for their distributions
and then added to the real data of the SDSS. The cluster detection
algorithm is applied to the combined data to find the input mock
clusters. The detection rate of mock clusters is considered as the
indication of completeness of our detection procedure. 

First, we generate a population of halos with masses
$M_{200}\ge0.6\times10^{14}~M_{\odot}$ following a mass function of
\citet{jfw+01} in a cosmology with $\Omega_m=0.3$ and $\sigma_8=0.9$
in the redshift range of $0.05<z<0.60$.
Then, we relate the cluster mass to cluster richness (i.e., total
luminosity in units of $L^{\ast}$) by
\begin{equation}
\log M_{200}=-1.49+1.17\log(C\times R_{L\ast,\rm in}),
\label{m200s}
\end{equation}
where $C$ is a scaling factor between input richness $R_{L\ast,\rm
  in}$ and output richness $R_{L\ast,\rm out}$ of our detection
procedure. The discrepancy between $R_{L\ast,\rm in}$ and
$R_{L\ast,\rm out}$ is caused by the uncertainty of galaxy
photo-$z$. Because the Equation~(\ref{m200}) is obtained for the
estimated richness based on the SDSS data, we adjust this factor $C$
to make output richness consistent with Equation~(\ref{m200}). We
first set $C=1$ for Equation~(\ref{m200s}) to get input
richness. Member galaxies are simulated for each halo within a radius
of $r_{200}$. The luminosity function and density profile of member
galaxies are taken as described in \citet{whl09}. Here, the
luminosities of member galaxies are evolution-corrected, as mentioned
in Section~\ref{scaling}. We assume that the uncertainty of
photometric redshift of member galaxies follows a Gaussian probability
function with a standard deviation of $\sigma_z$, but varies with
redshift in the form of $\sigma_z=0.03(1+z)$.

We add the mock clusters into the real SDSS data and apply our cluster
detection algorithm to measure the output
richness. Figure~\ref{outrich} shows the correlation between the input
richness and the output richness.  The best fit gives
\begin{equation}
R_{L\ast,\rm out}=(-0.52\pm0.16)+(0.81\pm0.01)R_{L\ast,\rm in}.
\end{equation}
Therefore, we get the factor $C=1/0.81$ for Equation~(\ref{m200s}) so
that the output richness $R_{L\ast,\rm out}$ can be related to cluster
mass by Equation~(\ref{m200}). 

Subsequently, we use Equation~(\ref{m200s}) to get a new population of
input richness. Again, we apply our method to detect the mock
clusters.  A cluster is detected if the output richness
$R_{L\ast,\rm out}\ge 12$ and $N_{200}\ge8$. The detection rate is
given to be the number of detected mock clusters divided by the total
number of added mock clusters. Figure~\ref{detrate} shows the
detection rate as a function of redshift. The detection rate is nearly
100\% for clusters with masses $M_{200}>2\times10^{14}~M_{\odot}$
(richness $R_{L\ast,\rm out}>34$) up to redshift of $z\sim0.5$, and
$\sim$75\% for clusters of $M_{200}>0.6\times10^{14}~M_{\odot}$
(richness $R_{L\ast,\rm out}>12$) up to redshift of $z\sim0.42$.  The
detection rate drops at higher redshifts, because the member galaxy
selection of our algorithm is consistent up to $z\sim0.42$, as
mentioned in Section~\ref{algorithm}.  In Figure~\ref{detrate}, we
also show the detection rate as a function of cluster mass for
clusters of $z<0.42$. More than 95\% of clusters of
$M_{200}>1\times10^{14}~M_{\odot}$ are detected.

\subsection{False detection rate}
\label{false}

\begin{figure}
\epsscale{1.}
\plotone{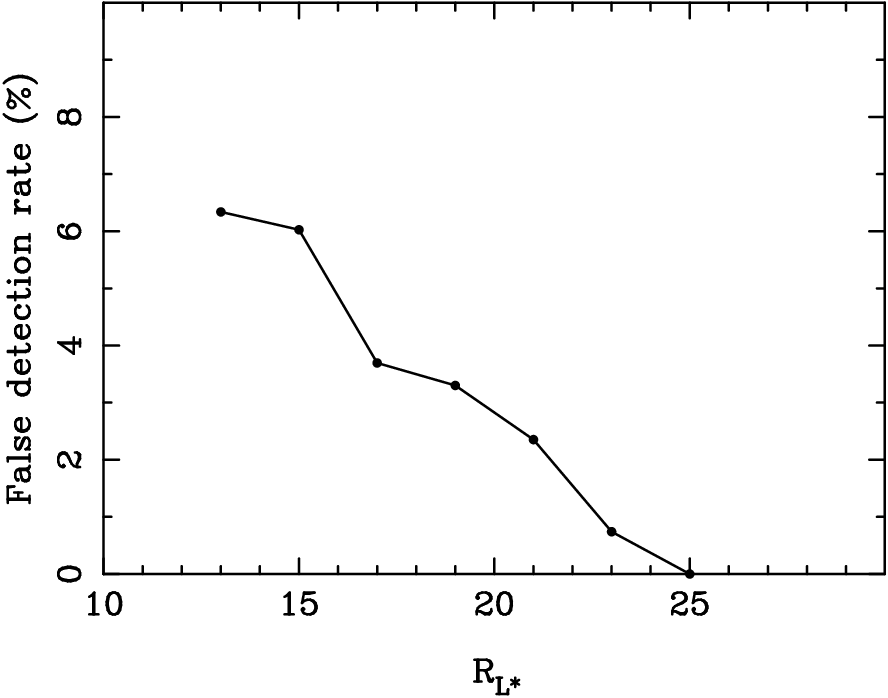}
\caption{False detection rate as a function of cluster richness.
\label{falrich}}
\end{figure}

The presence of the large-scale structures makes it possible to detect
false clusters because of projection effect. We perform a Monte Carlo
simulation with the real SDSS data to estimate the false detection
rate. First, we discard the member candidates of identified clusters
within a radius of $r_{200}$ and a photo-$z$ gap of $z\pm
0.04(1+z)$. Second, we shuffle the data following \citet{wh11}.  Our
new cluster detection algorithm is applied to the shuffled data.
After above procedures, we get the ``false clusters'' which satisfy
the criteria of $R_{L\ast}\ge12$ and $N_{200}\ge8$. The false
detection rate is calculated to be the total number of ``false
clusters'' identified from the shuffled data divided by the number of
clusters identified from the original SDSS data. We show that the
false detection rate is 6.0\% for a cluster richness of 12, but
decreases to less than 1\% for a richness of 23 (see
Figure~\ref{falrich}).

Note that we have removed some obvious contaminations from cluster
list in the cluster identification (Section~\ref{algorithm}). Here,
most of the contaminations cannot be found when shuffling the
cataloged galaxy data. Hence, removing the contaminations slightly
decreases the false detection rate.

\section{Comparison with previous cluster samples}

We compare our cluster sample in this paper (hereafter WHL12) with
previous ones, including the classical Abell sample from the Palomar
Sky Survey \citep{aco89}, the maxBCG \citep{kma+07b}, WHL09
\citep{whl09}, GMBCG \citep{hmk+10} and AMF \citep{spd+11} cluster
samples from the SDSS. We also compare our clusters with X-ray and SZ
cluster samples.

\subsection{Comparison with the Abell clusters}

We get the Abell cluster sample \citep{aco89} from the NASA/IPAC
Extragalactic Database. For the Abell clusters without redshift
measurements, we take their redshifts from the SDSS photo-$z$
data. There are 1844 Abell clusters with redshifts $z>0.05$ in the sky
coverage of SDSS-III, of which 1688 (92\%) clusters are within a
projected separation of $r_{200}$ and redshift difference of $|\Delta
z|\le0.05$ (about 2\,$\sigma$ of the uncertainty of the redshift
difference) from the WHL12 clusters. The non-matched Abell clusters
are relatively poor clusters or projected to neighbors of the large
scale structure \citep{whl09}. We mark the Abell clusters in Column
10 of Table~\ref{cat}.

\subsection{Comparison with the maxBCG clusters}

\begin{figure}
\epsscale{1.}
\plotone{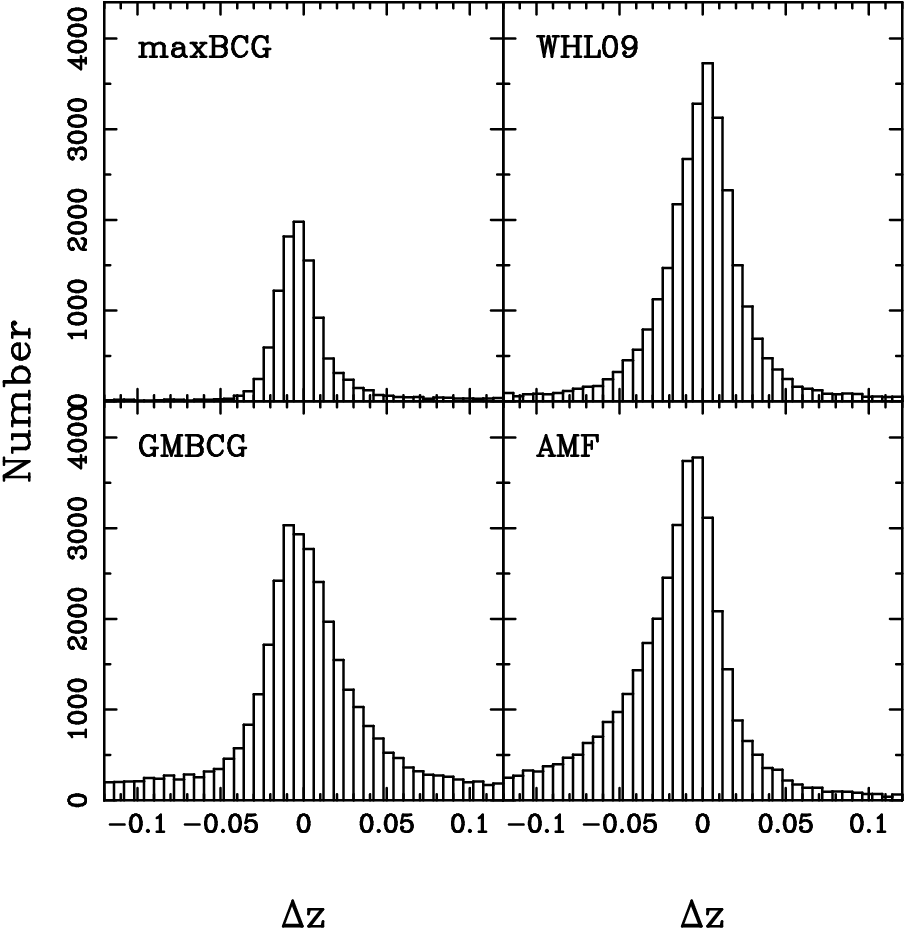}
\caption{Distribution of redshift differences between the WHL12 
clusters and the maxBCG, WHL09, GMBCG, and AMF clusters within a 
projected separation of $r_{200}$.
\label{hist_dz}}
\end{figure}

\begin{figure}
\epsscale{1.}
\plotone{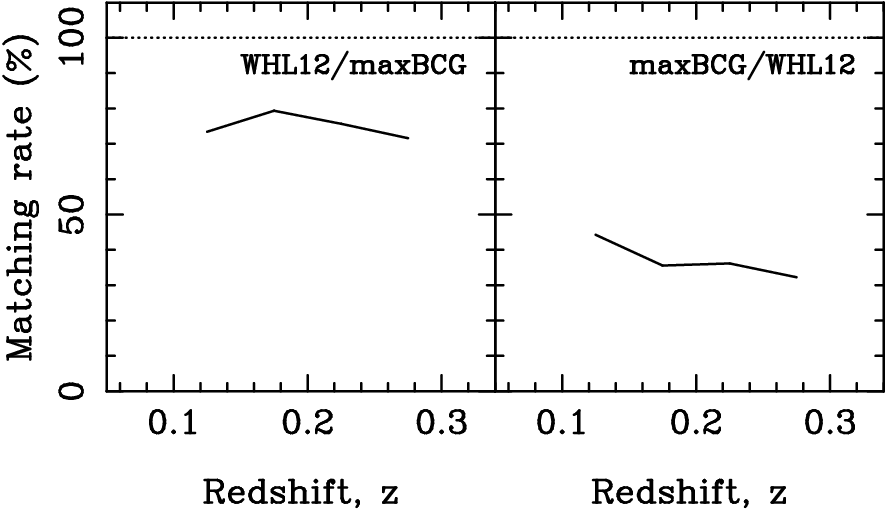}\\[2mm]
\plotone{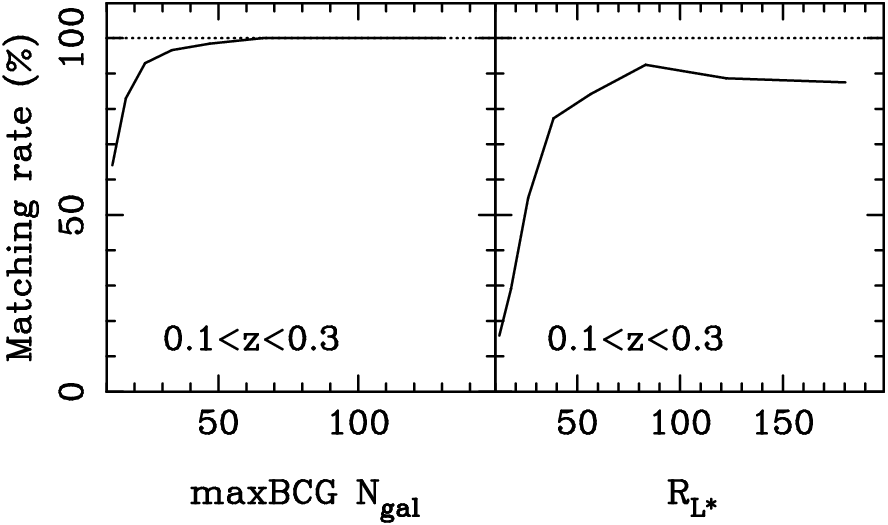}
\caption{Upper: matching rate between the WHL12 clusters and the
  maxBCG clusters, in the same sky coverage of the SDSS DR4, as a
  function of redshift. Lower: matching rate between the
  clusters of $0.1<z<0.3$ in the two samples as a function of cluster
  richness. The left panels show the matching rate of the WHL12
  clusters to the maxBCG clusters, and the right ones are the matching
  rate of the maxBCG clusters to the WHL12 clusters.
\label{rate_max}}
\end{figure}

\citet{kma+07b,kma+07a} developed a ``red-sequence cluster finder'',
maxBCG, to detect clusters dominated by red galaxies. They identify
clusters with a richness of $N_{\rm gal}\ge10$, where $N_{\rm gal}$ is the
number of galaxies brighter than $0.4L^{\ast}$ in the $i$ band within
$r_{200}$ and $2\,\sigma$ of the ridgeline colors. From the sky
coverage of the SDSS DR4, they obtained a complete volume-limited
sample containing 13,823 clusters in the redshift range of $0.1\le z\le
0.3$.  The maxBCG sample is approximately 85\% complete for the
clusters with masses $M>1\times10^{14}~M_{\odot}$.

We cross-match the WHL12 clusters with the maxBCG clusters in the sky
coverage of the SDSS DR4. Figure~\ref{hist_dz} shows the redshift
difference ($\Delta z=z_{\rm p}-z_{\rm maxBCG}$) between the WHL12
clusters and the maxBCG clusters (also the WHL09, GMBCG, and AMF
clusters for discussion below) within a projected separation of
$r_{200}$. It shows that a value of $|\Delta z|\le0.05$ is an
appropriate selection threshold for cluster matching in redshift
difference (top left panel of Figure~\ref{hist_dz}). We find
that 10,301 (75\%) of 13,823 maxBCG clusters are matched with the
WHL12 clusters within a separation of $r_{200}$ and a redshift
difference of $|\Delta z|\le0.05$. Figure~\ref{rate_max} shows the
matching rates between the WHL12 clusters and the maxBCG clusters as a
function of redshift. We find that 70\%--80\% maxBCG clusters are
matched with the WHL12 clusters in the redshift range of $0.1\le z\le
0.3$, while only 30\%--40\% of the WHL12 clusters are matched with the
maxBCG clusters because our sample is more complete for poor
clusters. The matching rates increase with cluster richness for both
samples. As shown in Figure~\ref{rate_max}, 95\% of the maxBCG
clusters of $N_{\rm gal}\ge20$ are matched with the WHL12 clusters,
and 85\% of the WHL12 clusters of $R_{L\ast}\ge40$ are matched with
the maxBCG clusters.
Those non-matched maxBCG clusters mostly have a richness smaller than
the criteria of our cluster finding algorithm. A small number of the
maxBCG clusters of $N_{\rm gal}\ge20$ are missing in our sample. As we
have checked, some of them are merged in the friend-of-friend
algorithm of our procedure (steps 2 and 5), and some of them can be found
with a larger separation. If we match the clusters within a separation
of $1.5\,r_{200}$, $97\%$ of the maxBCG clusters of $N_{\rm gal}\ge20$
can be found in our sample.

\begin{figure}[t!]
\epsscale{1.}
\plotone{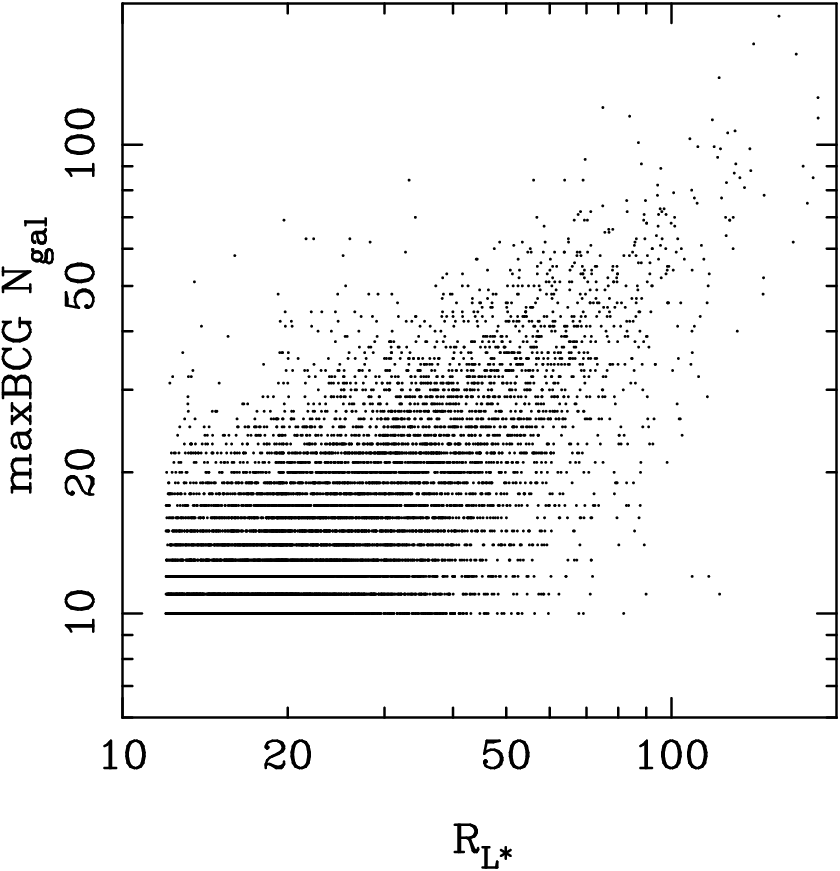}
\caption{Comparison of richness between $N_{\rm gal}$ in the maxBCG and 
$R_{L\ast}$ in this paper for the matched clusters. 
\label{rich_max}}
\end{figure}

Figure~\ref{rich_max} shows the correlation between the maxBCG
richness, $N_{\rm gal}$, and the richness of this paper, $R_{L\ast}$,
for the matched clusters.

\subsection{Comparison with the WHL09 clusters}

From the SDSS DR6, \citet{whl09} identified 39,668 clusters of
galaxies in the redshift range $0.05\le z<0.6$. Monte Carlo
simulations show that the sample is complete up to redshift $z=0.42$
for clusters with a richness $R\ge16.7$. The cluster richness is the
number of member candidates of $M_r\le-21$ within a radius of 1 Mpc
and a photo-$z$ of $\pm0.04(1+z)$ after background subtraction.

Comparing to \citet{whl09}, we modify the richness estimate and
cluster identification criteria. A cluster is identified in this paper
when the richness $R_{L\ast}\ge 12$ and the number of member
candidates $N_{200}\ge 8$ within $r_{200}$, which is looser than the
criteria of $N_{\rm gal}\ge 8$ within a radius of 0.5 Mpc in
\citet{whl09}. The cluster sample in this paper is more complete for
clusters of low richness.  The completeness is 40\% for clusters with
a mass of $10^{14}~M_{\odot}$ (richness of $\sim$12) in \citet{whl09},
while it is $\sim$90\% in this paper.

\begin{figure}[t]
\epsscale{1.}
\plotone{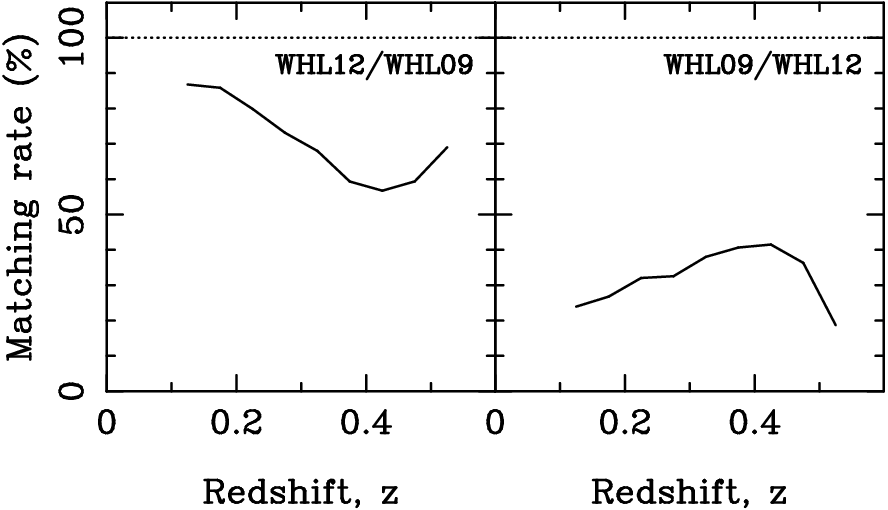}\\[2mm]
\plotone{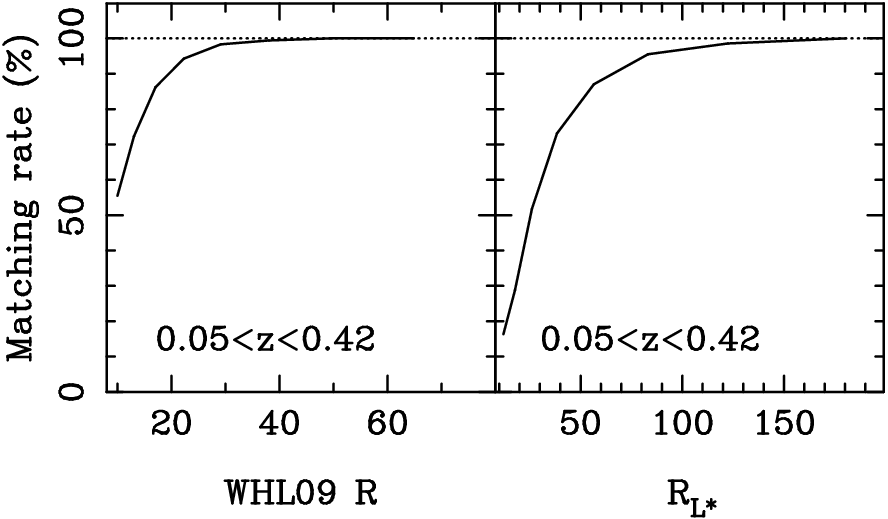}
\caption{Same as Figure~\ref{rate_max} but between the WHL09 clusters and 
the WHL12 clusters in the sky coverage of the SDSS DR6.
\label{rate_whl}}
\end{figure}

\begin{figure}
\epsscale{1.}
\plotone{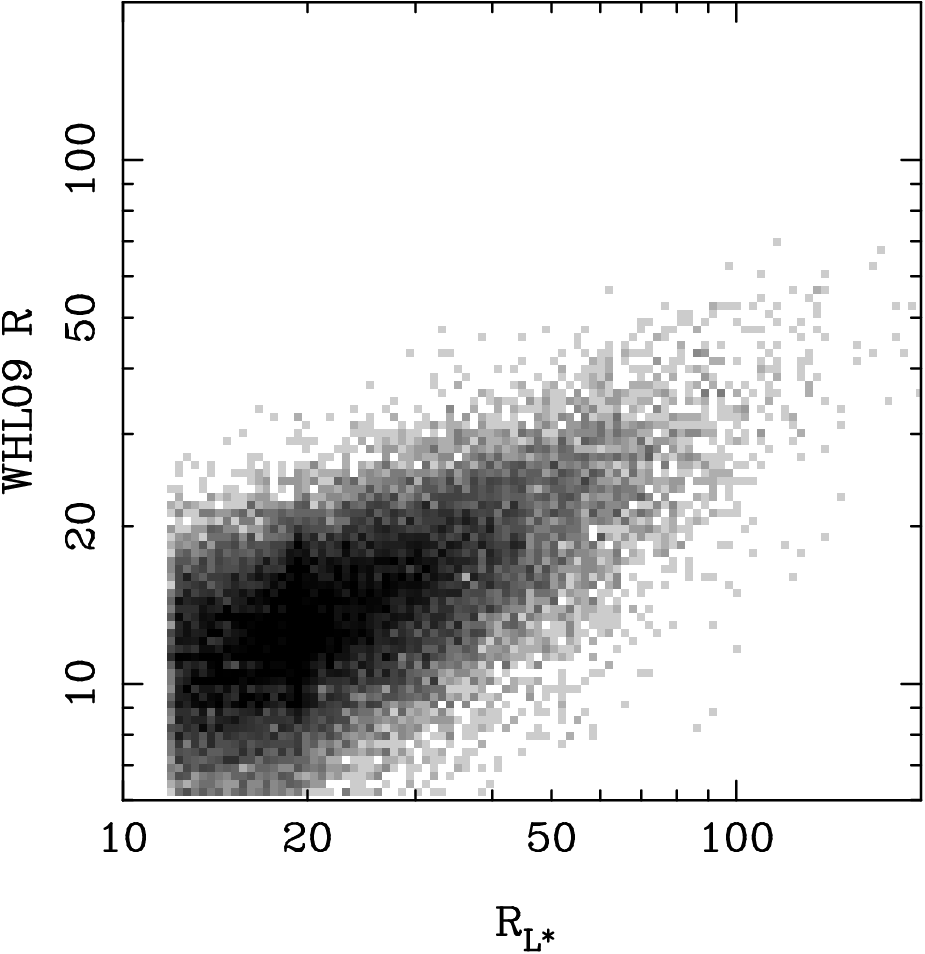}
\caption{Same as Figure~\ref{rich_max} but between $R$ in WHL09 and 
$R_{L\ast}$ in this paper.
\label{rich_whl}}
\end{figure}

Matching is applied between two cluster samples in sky coverage of the
SDSS DR6. We find that 25,929 (65\%) of 39,668 WHL09 clusters are
matched with the WHL12 clusters within a separation of $r_{200}$ and
a redshift difference of $|\Delta z|\le0.05$.  Figure~\ref{rate_whl} shows
the matching rate as a function of redshift between the WHL12 clusters
and WHL09 clusters. The matching rates of the WHL12 clusters to WHL09
clusters decrease from 85\% at $z\sim0.1$ to 55\% at $z\sim0.4$. The
matching rates of the WHL09 clusters to the WHL12 clusters increase
from 20\% at $z\sim0.1$ to 40\% at $z\sim0.4$. The matching rates vary
as a function of cluster richness within $0.05\le z<0.42$, 90\% of the
WHL09 clusters of $R\ge15$ are matched with the WHL12 clusters, 86\%
of the WHL12 clusters of $R_{L\ast}\ge40$ are matched with the WHL09
clusters. If we match the clusters within a separation of
$1.5\,r_{200}$, $91\%$ of the WHL09 clusters of $R\ge15$ are matched.
The non-matched WHL09 clusters are probably merged in the
friend-of-friend algorithm of our procedure, or are poorer than the
criteria of cluster finding in this work.

Figure~\ref{rich_whl} shows a tight correlation between the WHL09
richness, $R$, and the richness of this paper, $R_{L\ast}$, for the
matched clusters.

\subsection{Comparison with the GMBCG clusters}

\begin{figure}
\epsscale{1.}
\plotone{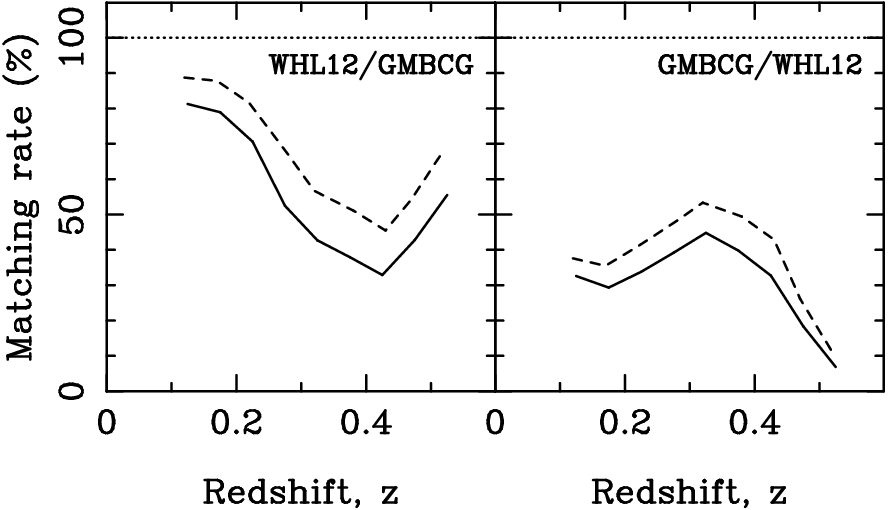}\\[2mm]
\plotone{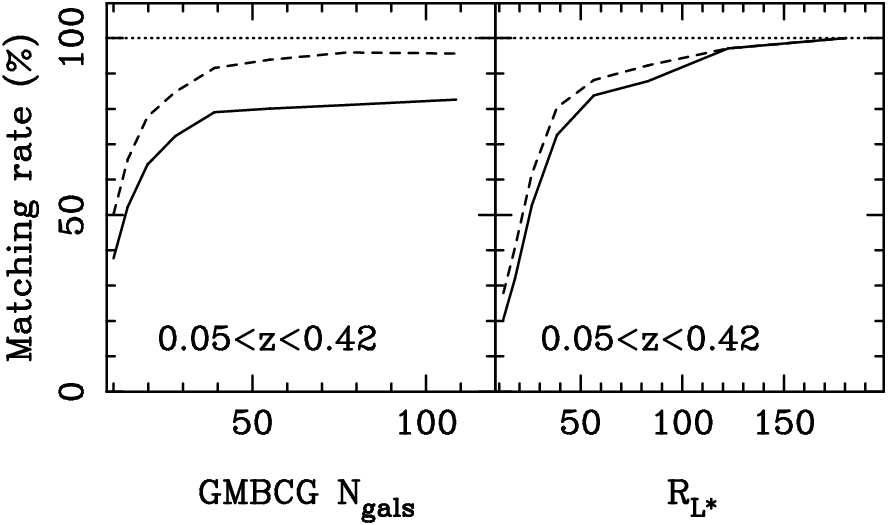}
\caption{Same as Figure~\ref{rate_whl} but between the GMBCG clusters
  and the WHL12 clusters in the sky coverage of the SDSS DR7. The
  solid lines represent the matching rate within a separation of
  $r_{200}$ and a redshift difference of $|\Delta z|\le0.05$.The
  dashed lines represent the matching rate within a separation of
  $1.5\,r_{200}$ and a redshift difference of $|\Delta z|<0.1$.
\label{rate_gmbcg}}
\end{figure}

\begin{figure}
\epsscale{1.}
\plotone{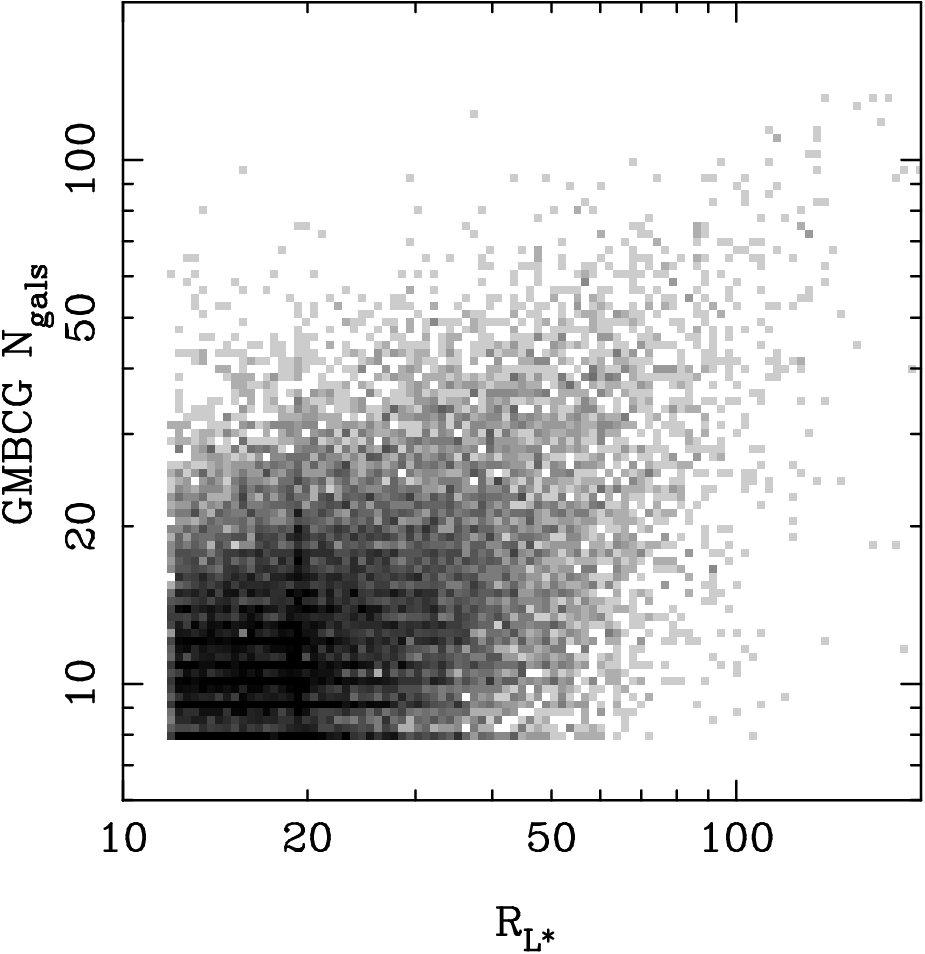}
\caption{Same as Figure~\ref{rich_max} but between $N_{\rm gal}$ in 
the GMBCG and $R_{L\ast}$ in this paper.
\label{rich_gmbcg}}
\end{figure}

\citet{hmk+10} presented a GMBCG algorithm, which detects clusters by
identifying the red sequence and the BCG feature. From the SDSS DR7,
they found 55,424 clusters in the redshift range of $0.1<z<0.55$. They
provide two richnesses, the weighted richness which is the total
number of galaxies brighter than $0.4L^{\ast}$ within $r_{200}$
weighted by a factor from the fitting of color distribution, and the
scaled richness which is the number of galaxies within $r_{200}$ and
$2\,\sigma$ of the ridgeline colors. We adopt the weighted richness as
GMBCG richness $N_{\rm gals}$ if available, otherwise the scaled
richness.

We cross-match the WHL12 clusters with the GMBCG clusters in the sky
coverage of the SDSS DR7. We find that 25,370 (46\%) of 55,424 GMBCG
clusters are matched with the WHL12 clusters within a separation of
$r_{200}$ and a redshift difference of $|\Delta z|\le0.05$ (see
Figure~\ref{rate_gmbcg}). The matching rates of the WHL12 clusters to
the GMBCG clusters decrease from 80\% at $z\sim0.1$ to 40\% at
$z\sim0.4$. The matching rates of the GMBCG clusters to the WHL12
clusters are 30\%--40\% within $z<0.4$. Within the redshift range of
$0.05\le z<0.42$, the matching rates increase with cluster richness
for both samples. We find that 72\% of the GMBCG clusters of $N_{\rm
  gals}\ge20$ are matched with the WHL12 clusters, 83\% of the WHL12
clusters of $R_{L\ast}\ge40$ are matched with the GMBCG clusters. The
fraction of missing rich GMBCG clusters is larger than those of the
maxBCG and WHL09 clusters. This may be due to a broader distribution
of redshift difference (see Figure~\ref{hist_dz}). If we match
clusters within a separation of $1.5\,r_{200}$ and a redshift difference
of $|\Delta z|<0.1$, 55\% of the GMBCG clusters are matched with the
WHL12 clusters, 85\% of the GMBCG rich clusters of $N_{\rm gals}\ge20$
are matched (see dashed lines in Figure~\ref{rate_gmbcg}).

We note that a few rich GMBCG clusters are still missing. Some of them
are merged in the friend-of-friend algorithm of our procedure, or have
a redshift difference of $|\Delta z|>0.1$ from the WHL12
clusters. Most of other missing GMBCG clusters are poorer than the
criteria of cluster finding in this work. We note a large scatter
between the GMBCG richness $N_{\rm gals}$ and $R_{L\ast}$ (see
Figure~\ref{rich_gmbcg}).

\subsection{Comparison with the AMF clusters}

\begin{figure}
\epsscale{1.}
\plotone{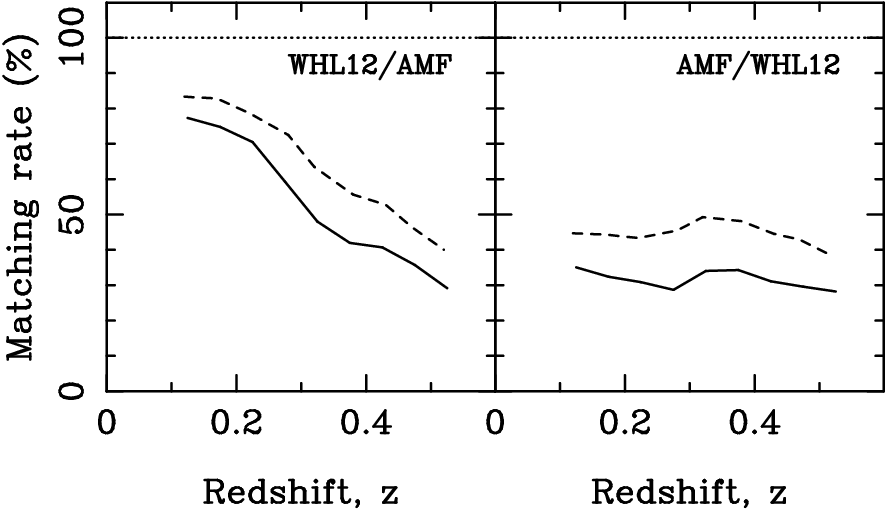}\\[2mm]
\plotone{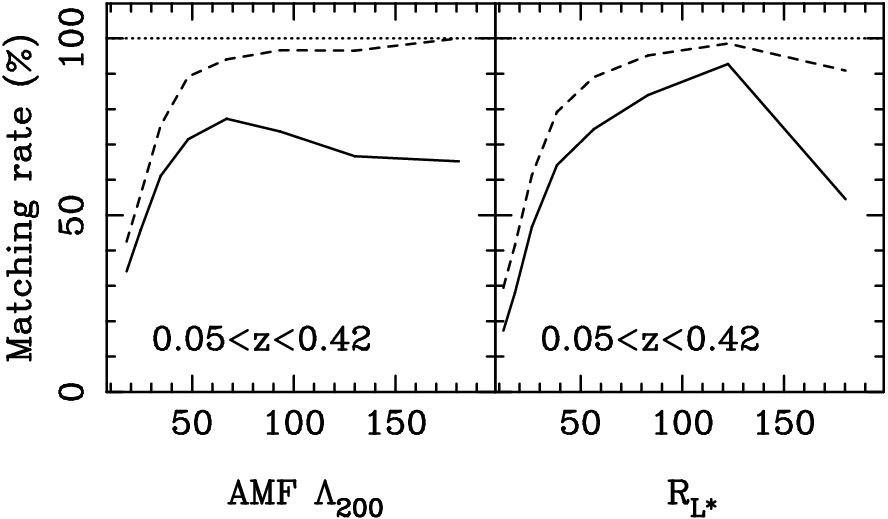}
\caption{Same as Figure~\ref{rate_gmbcg} but between the AMF clusters and 
the WHL12 clusters in the sky coverage of the SDSS DR6.
\label{rate_amf}}
\end{figure}

\begin{figure}
\epsscale{1.}
\plotone{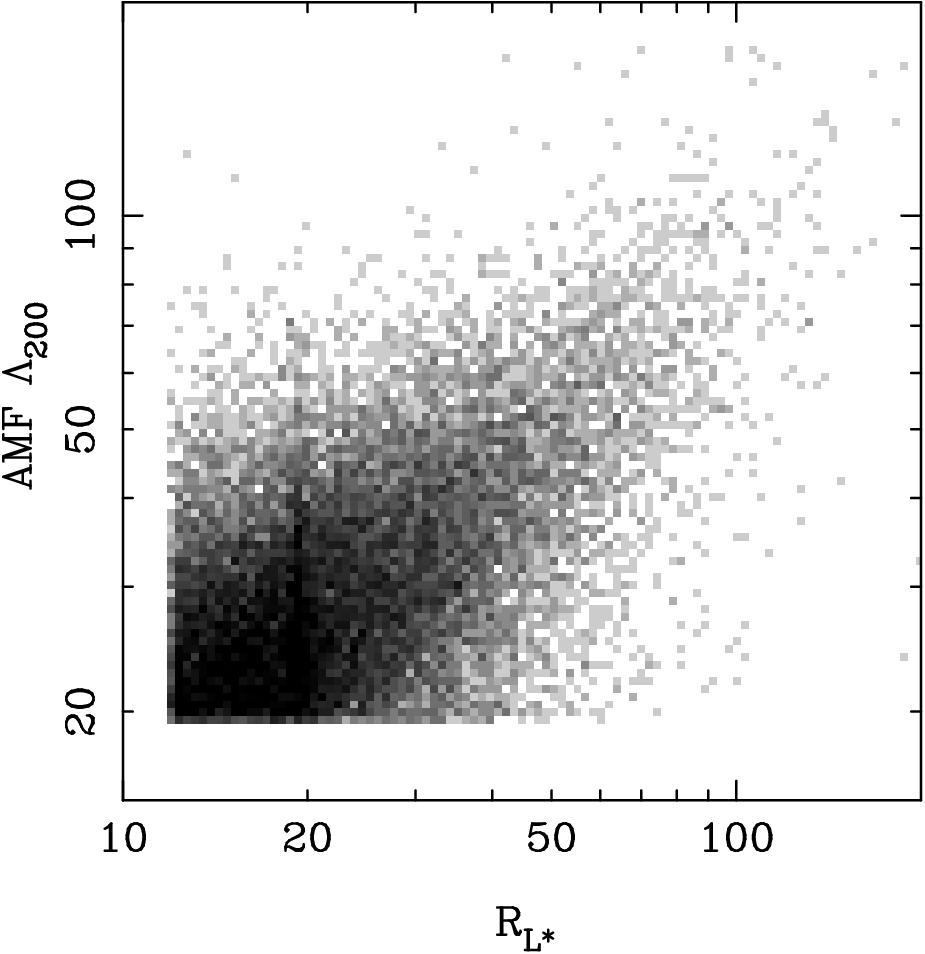}
\caption{Same as Figure~\ref{rich_max} but between $\Lambda_{200}$ in 
the AMF and $R_{L\ast}$ in this paper.
\label{rich_amf}}
\end{figure}

\citet{spd+11} used a modified AMF algorithm to identify clusters
using photometric redshifts of galaxies. From the SDSS DR6, 69,173
clusters are identified in the redshift range $0.045<z<0.78$ with a
richness of $\Lambda_{200}\ge20$. The richness of the AMF clusters is
defined as the total luminosity within $r_{200}$ in units of
$L^{\ast}$, the same as that in this work.

We cross-match the WHL12 clusters with the AMF clusters in the sky
coverage of the SDSS DR6. Within a separation of $r_{200}$ and a redshift
difference of $|\Delta z|\le0.05$, 27,966 (40\%) of 69,173 AMF clusters
are matched with the WHL12 clusters. Figure~\ref{rate_amf} shows the
matching rate as a function of redshift. The matching rates of the
WHL12 clusters to the AMF clusters decrease from 80\% at $z\sim0.1$ to
30\% at $z\sim0.5$. The matching rates of the AMF clusters to the
WHL12 clusters are constantly $\sim$30\%. The matching rates vary with
cluster richness within $0.05\le z<0.42$, 72\% of the AMF clusters of
$\Lambda_{200}\ge40$ are matched with the WHL12 clusters, 74\% of the
WHL12 clusters of $R_{L\ast}\ge40$ are matched with the AMF
clusters. We note that many AMF clusters have a large redshift
difference of $|\Delta z|>0.05$ (see Figure~\ref{hist_dz}). If we
match clusters within a separation of $1.5\,r_{200}$ and a redshift
difference of $|\Delta z|<0.1$, 91\% of the AMF rich clusters of
$\Lambda_{200}\ge40$ are matched with the WHL12 clusters (see dashed
lines in Figure~\ref{rate_amf}). Those non-matched AMF clusters are
missing due to friend-of-friend merging of our algorithm, large
redshift difference from the WHL12 clusters and large scatter between
two richness estimates.

Figure~\ref{rich_amf} shows the correlation between the AMF richness
$\Lambda_{200}$ and the $R_{L\ast}$ for the matched clusters.

\subsection{Correlations with X-Ray Measurements}
\label{cluster_xray}

\begin{figure}
\epsscale{1.1}
\plotone{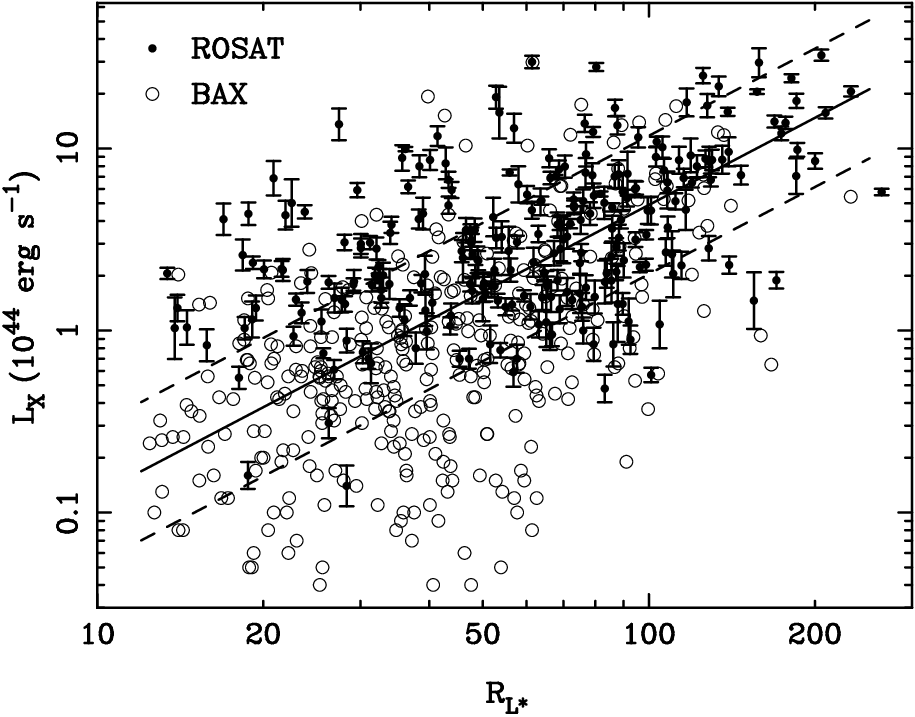}\\[2mm]
\plotone{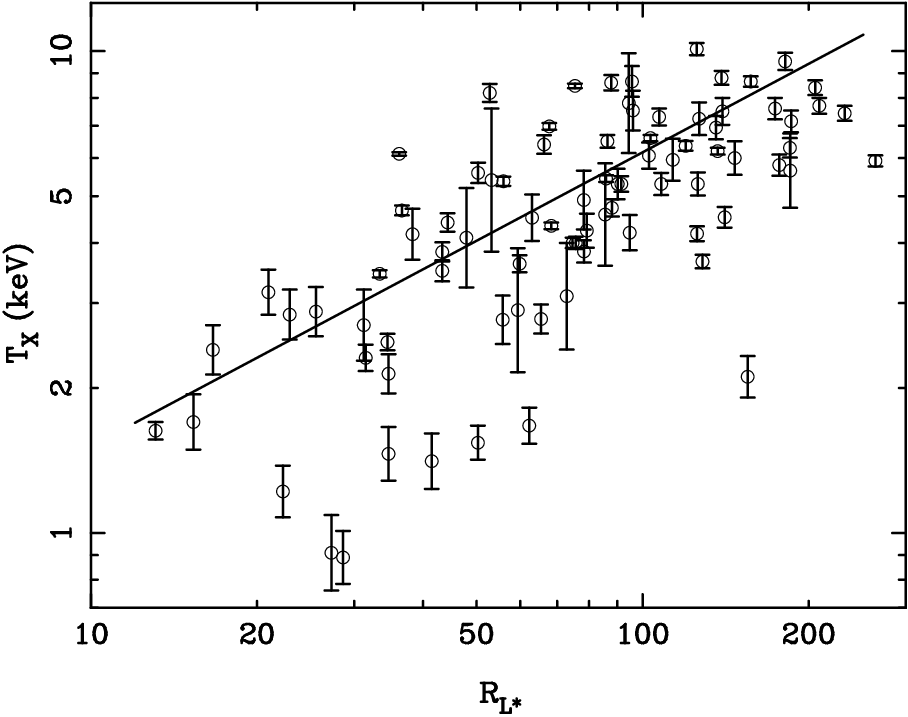}
\caption{Correlations between the cluster richness $R_{L\ast}$ with
  the X-ray luminosity $L_X$ (the upper panel) and temperature $T_X$
  (the bottom panel).  The solid lines are the best fits for the
  correlations. The dashed lines in the upper panel show the
  $\pm\sigma_{\log L_X}$ to the best-fit relation.
\label{rich_xray}}
\end{figure}

With luminous member galaxies well discriminated in the identification
procedure, good correlations between the cluster richness with X-ray
measurements are expected \citep{whl09}.

We get X-ray clusters from the Northern ROSAT All-Sky (NORAS)
Galaxy Cluster Survey \citep{bvh+00} and the ROSAT-ESO Flux Limited
X-ray (REFLEX) Galaxy cluster survey \citep{bsg+04}. The NORAS sample
contains 378 clusters and the REFLEX contains 447 clusters. The {\it ROSAT}
survey provides X-ray luminosity for detected clusters. We also get
X-ray clusters from the query of the X-ray Cluster Database
(BAX\footnote{http://bax.ast.obs-mip.fr/}), where luminosities for
1028 clusters and temperatures for 210 clusters are available compiled
from reference. The {\it ROSAT} clusters mostly are massive with a
luminosity greater than $10^{44}~{\rm erg~s^{-1}}$ in the 0.1--2.4 keV
band, while the BAX sample contains many less massive clusters with a
low luminosity less than $10^{44}~{\rm erg~s^{-1}}$. Previous work
shows that the luminosity in both samples is statistically consistent
\citep{spd+11}. Here, we combine two sample for the correlation with
our cluster richness. For the clusters in both {\it ROSAT} and BAX samples,
we adopt the X-ray luminosity from {\it ROSAT} data.

X-ray clusters are matched with the WHL12 clusters within a separation
of $r_{200}$ and a redshift difference of $|\Delta z|\le0.05$. We find
611 matched X-ray clusters with luminosities measured, 84 matched
X-ray clusters with temperatures measured. Figure~\ref{rich_xray}
shows the correlations between cluster richness with X-ray luminosity
and temperature for the matched clusters. The best fits give
\begin{equation}
\log L_{X,44}=(-2.49\pm0.16)+(1.59\pm0.09)\log R_{L\ast},
\label{lxr}
\end{equation}
and
\begin{equation}
\log T_X=(-0.43\pm0.05)+(0.61\pm0.05)\log R_{L\ast},
\end{equation}
where $L_{X,44}$ refers to X-ray luminosity in the 0.1--2.4 keV band
in units of $10^{44}~{\rm erg~s^{-1}}$, $T_X$ refers to X-ray
temperature in units of keV. The tight correlations suggest that the
richness we estimate is reasonable and statistically reliable. The
slopes of the $L_X$--$R_{L\ast}$ and the $T_X$--$R_{L\ast}$
relations are in agreement with those of \citet{spd+11}. We measure
the scatter in X-ray luminosity to the best-fit relation and get a
value of $\sigma_{\log L_X}=0.38$, i.e., $\sigma_{\ln L_X}=0.87$.  The
scatter is in agreement with that for the correlation of $L_X$ with
the maxBCG richness \citep{rmb+08}, but larger than those for the
correlations with the improved maxBCG richness \citep{rrk+09,rkr+12}.
For a similar comparison, we measure the scatter in cluster mass to
the best-fit $M_{200}$--$R_{L\ast}$ relation given in
Section~\ref{scaling} and get $\sigma_{\log M_{200}}=0.21$, i.e.,
$\sigma_{\ln M_{200}}=0.48$. The scatter in mass based on our richness
is also in agreement with that based on the maxBCG richness
\citep{rre+09}.

\subsection{Correlations with the SZ Measurements}
\begin{figure}
\epsscale{1.2}
\plotone{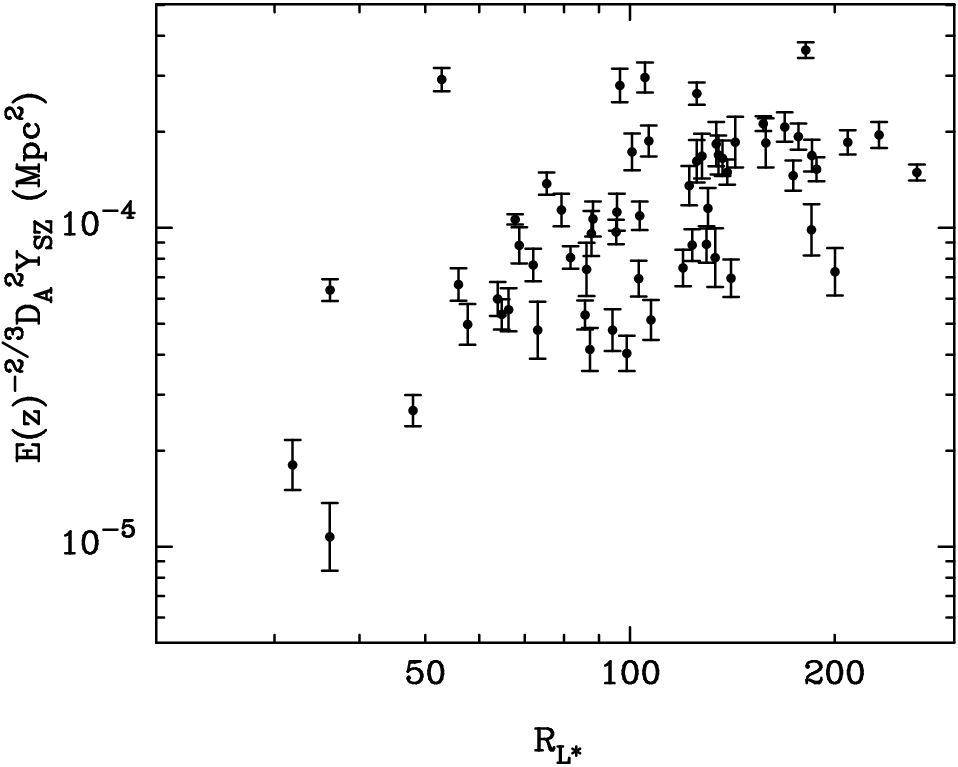}
\caption{Correlations between the cluster richness 
with the SZ measurement by Planck.
\label{rich_sz}}
\end{figure}

The SZ effect is the result of high energy electrons interacting with
the cosmic microwave background radiation through inverse Compton
scattering. It has been detected around rich galaxy clusters due to a
high temperature of host ionized gas \citep{cjg96,cjg+00}. The SZ
signal is characterized by the quantity $D^2_{A(z)} Y_{\rm SZ} =
(\sigma_T/m_ec^2)\int PdV$, where $D_{A(z)}$ is the angular distance
to a cluster at redshift $z$, $\sigma_T$ is the Thomson cross-section,
$c$ is the speed of light, $m_e$ is the electron rest mass, $\int P
dV= \int n_ekT_e dV$ is the integration of the pressure of hot ionized
gas over the cluster volume. As the pressure is related to
gravitational potential, $D^2_{A(z)} Y_{\rm SZ}$ is expected to be a
proxy of cluster mass \citep{bjl+08}.

Recently, the Planck releases an SZ cluster sample including 189
clusters \citep{planck11}. We find that 71 SZ clusters are located at
the sky coverage of the SDSS-III. Six of them have redshifts less than
0.05. Other 65 clusters are matched with the WHL12 clusters, of which
61 have X-ray measurements.
Figure~\ref{rich_sz} shows the correlations between the richness and
the SZ measurements by the Planck for the 61 clusters, where
$Y_{\rm SZ}$ is the SZ signal measured within a radius of $5\,r_{500}$
and $E(z)=\sqrt{\Omega_{\Lambda}+\Omega_m(1+z)^3}$. The matched
clusters mostly have a richness of $R_{L\ast}> 50$. Clearly, there is
a positive correlation between cluster richness and SZ measurement.

\section{BCGs and LRGs in the clusters}

BCGs are luminous and usually located at the centers of
clusters. Their properties provide the information of galaxy formation
in the extreme dense environment. Using our sample, we can study the
evolution of the BCGs. Moreover, we cross-identity the BCGs with the
SDSS LRG sample.

\subsection{Magnitude evolution of the BCGs}

\begin{figure}
\epsscale{1.1}
\plotone{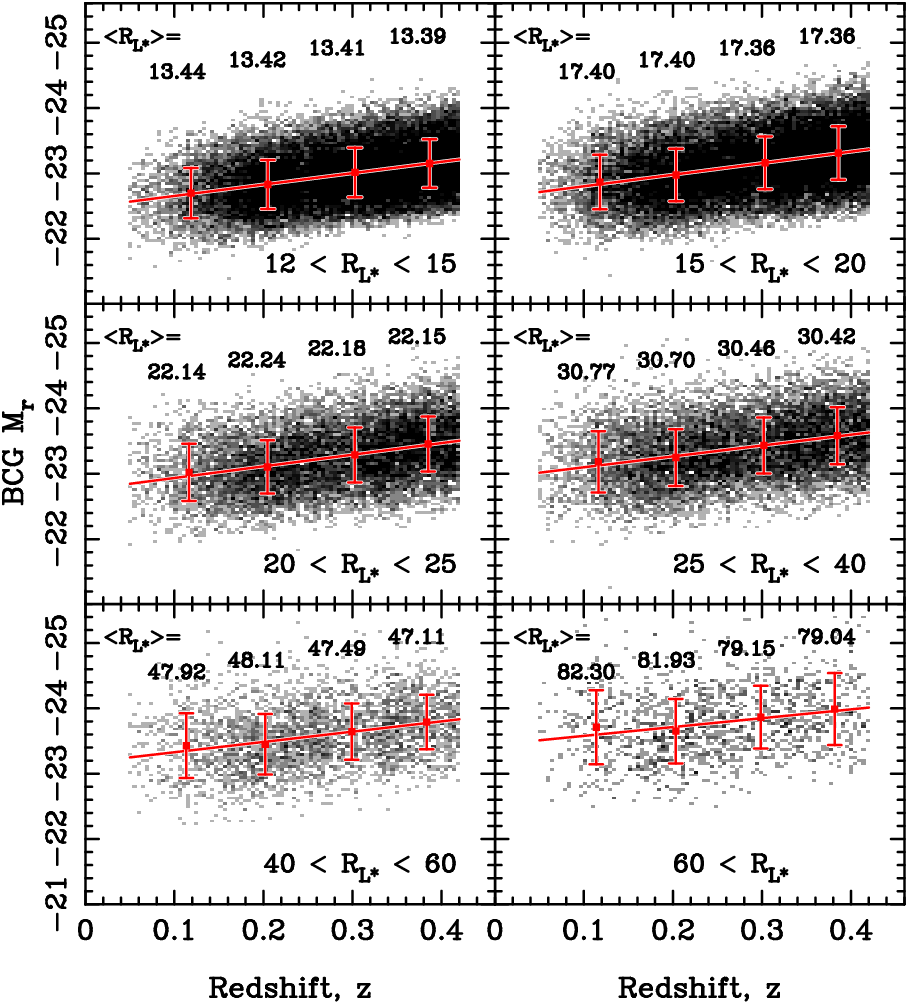}\\[2mm]
\plotone{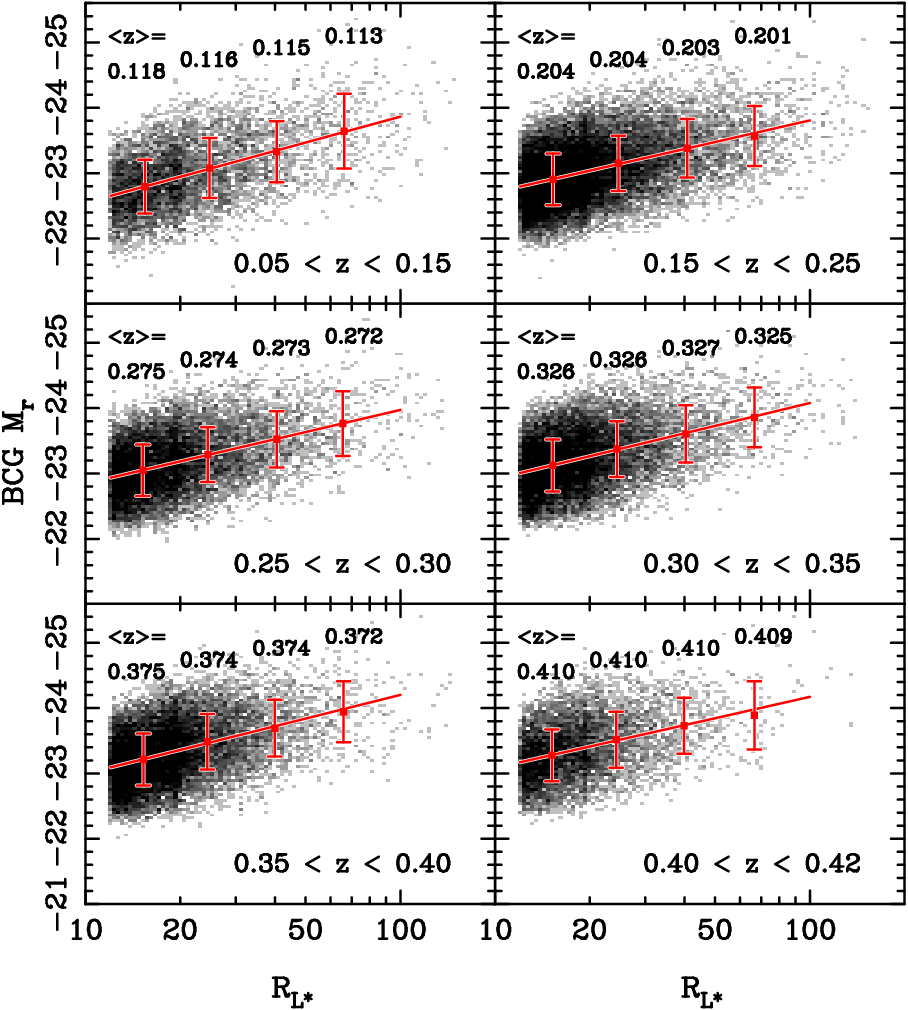}\\[2mm]
\caption{Upper: evolution of BCG absolute magnitude with cluster
  redshift in the range of $0.05\le z<0.42$ for six richness
  bins. The average absolute magnitudes and the dispersions are
    plotted for four redshift ranges. The average richnesses are
    calculated and marked for these redshift ranges. The solid lines
    show the best-fitting. Lower: correlation between BCG
  absolute magnitude with cluster richness for six redshift bins. 
Similarly, the average absolute magnitudes and the dispersions are
    plotted for four richness ranges, and the average redshifts are
    marked for these ranges in the plots.\\ 
(A color version of this figure is available in
  the online journal.)
\label{bcg1}}
\end{figure}

The BCGs in rich clusters are formed at redshift $z>2$ and evolved
passively \citep[e.g.,][]{ses+08}. The properties of BCGs are related
to their host clusters \citep{wh11}. Richer clusters tend to have
brighter BCGs. Here, we use our sample to address the magnitude
evolution of the BCGs with redshift and richness.

Figure~\ref{bcg1} shows the evolution of absolute magnitudes of the
BCGs with cluster redshift in the range of $0.05\le z<0.42$ for six
richness bins and with cluster richness for six redshift bins. We get
the mean values and fit the correlation (solid lines) for each
bin. Here, the variations of completeness and false detection rate are
considered by a weighting factor in the fitting. The weight is
calculated as the percentage of true cluster (i.e., $100\%-$false
detection rate) divided by completeness for each richness and redshift
bin. There are very small changes of the average richnesses within the
redshift ranges (upper panels) or the average redshifts within the
richness ranges (lower panels), indicating that the changes of BCG
magnitudes have been well decoupled for redshift and richness. The
best fits between absolute magnitude and redshift give
\begin{eqnarray}
M_r=(-22.48\pm0.01)&&-(1.74\pm0.03)z \nonumber \\
&&{\rm for}~ 12<R_{L\ast}<15; \nonumber \\
M_r=(-22.63\pm0.01)&&-(1.76\pm0.03)z \nonumber \\
&&{\rm for}~ 15<R_{L\ast}<20; \nonumber \\
M_r=(-22.76\pm0.01)&&-(1.77\pm0.05)z \nonumber \\
&&{\rm for}~ 20<R_{L\ast}<25; \nonumber \\
M_r=(-22.93\pm0.01)&&-(1.66\pm0.05)z \nonumber \\
&&{\rm for}~ 25<R_{L\ast}<40; \nonumber \\
M_r=(-23.17\pm0.03)&&-(1.58\pm0.09)z \nonumber \\
&&{\rm for}~ 40<R_{L\ast}<60; \nonumber \\
M_r=(-23.44\pm0.04)&&-(1.35\pm0.17)z \nonumber \\
&&{\rm for}~ 60<R_{L\ast}.
\end{eqnarray}
Clearly, the BCG are brighter in clusters of higher redshifts. As the
stellar population in the BCGs was formed at redshift $z>2$ and
becomes old with comic time, the BCGs become fainter at lower
redshifts. To verify the evolution of BCG magnitude with redshift,
using the AMF clusters \citep{spd+11} and the brightest one of three
BCG candidates for each cluster, we get
\begin{equation}
M_{r,\rm AMF}=(-22.43\pm0.01)-(1.54\pm0.04)z,
\end{equation}
which is roughly consistent with the results derived from our cluster
sample. The BCG magnitude evolution is in agreement with that of
$L^{\ast}$ found by \citet{bhb+03}.
In addition, we find that BCGs are brighter in richer clusters,
consistent with conclusions in previous works
\citep{wh11,spd+11}. From the fitting, the BCGs have an average
absolute magnitude of $-22.48$ for cluster richness of
$R_{L\ast}=12$--15 at redshift $z=0$, but a brighter magnitude of
$-23.44$ for cluster richness of $R_{L\ast}>60$. This tendency can be
clearly shown from the correlation between BCG absolute magnitude with
cluster richness for six redshift bins (the lower panel of
Figure~\ref{bcg1}). The evolution of the BCGs magnitudes with
redshift and the dependence of richness can be formulated together
as,
\begin{eqnarray}
M_r=(&-&21.25\pm0.01)-(1.75\pm0.03)z\nonumber \\
     &-&(1.10\pm0.03)\log R_{L\ast}.
\end{eqnarray}

\subsection{Cross-identification between BCGs and LRGs}

\begin{figure}[t]
\epsscale{1.1}
\plotone{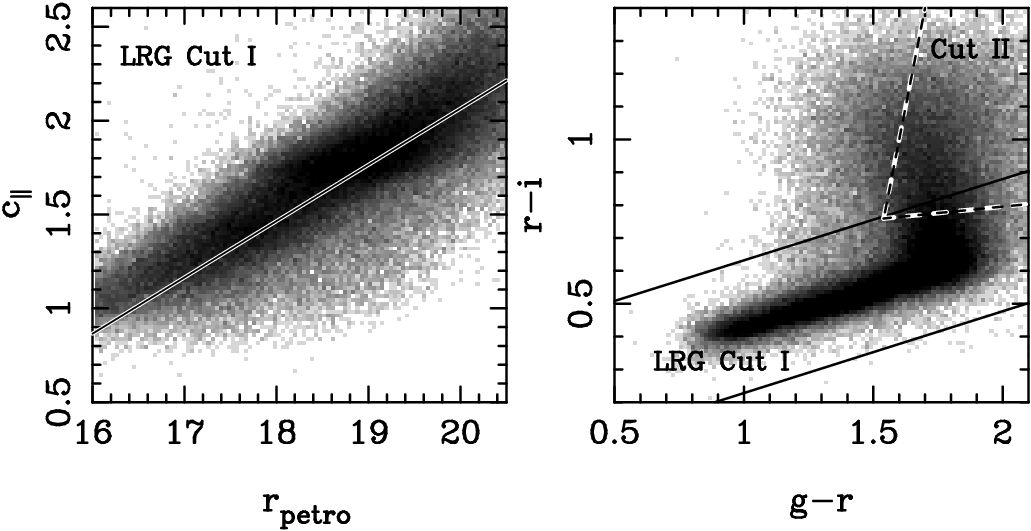}
\caption{Color--magnitude and color--color diagrams for BCGs in our 
clusters. The solid lines represent the color cuts of the 
SDSS LRG selection.
\label{sdsslrg}}
\end{figure}

The SDSS LRGs are selected to produce a volume-limited sample of
massive galaxies to redshift of 0.5. They are efficiently selected by
color cuts based on passive evolution model at redshifts $z>0.15$
\citep{eag+01}. Generally, LRGs are more massive and located at denser
environment than general galaxies. They are most likely BCGs or
brightest group galaxies. BCGs are luminous and mostly have red
colors. They are likely to be LRGs. However, this may not always be
true. \citet{cae+99} showed that about 27\% BCGs in X-ray bright
clusters have optical emission lines. They may have blue colors due to
recent star formation \citep{hmd10,wok+10,psp+11}. Therefore, many
BCGs are not LRGs.

First, we cross-identify the clusters in our sample with the SDSS
LRGs to address how many LRGs are member galaxies.  We get 112,191
LRGs of $z>0.15$ from the SDSS spectroscopic data.  Cross-matching
between LRGs with BCGs in our sample, we find that 28,336 (25\%) LRGs
are BCGs and 40,039 (36\%) LRGs are located within a projected
separation of $r_{200}$ and a redshift difference of 0.05. This suggests
that most LRGs are located not in our clusters but in an environment
poorer than our clusters.

On the other hand, we estimate the fraction of the BCGs as the SDSS
LRGs. Follow \citet{eag+01}, we match the colors of the BCGs with the
LRG selection criteria. Figure~\ref{sdsslrg} shows the color--magnitude
and color--color diagrams for the BCGs in our clusters. The color
$c_{\parallel}$ is defined as
\begin{equation}
c_{\parallel}=0.7(g-r)+1.2[(r-i)-0.177].
\end{equation}
If we only consider the color cuts and do not limit the Petrosian
magnitude $r_{\rm petro}$, the BCGs can be classified as LRGs for
those with the color $c_{\parallel}$ above the line in the left panel
of Figure~\ref{sdsslrg} and the colors, $g-r$ and $r-i$, between the
lines in the right panel. There are 83,740 (66\%) of 126,041 BCGs in
the redshift range $z>0.15$ satisfying the color cuts of the SDSS LRG
selection criteria. When the Petrosian magnitude limits of $r_{\rm
  petro}<19.2$ for Cut I and $r_{\rm petro}<19.5$ for Cut II are
considered, we find that 63\% of BCGs of $r_{\rm petro}<19.5$ are
LRGs.

\section{Summary}

We identify 132,684 clusters of galaxies in the redshift range
$0.05\le z<0.8$ using photometric redshifts of galaxies from the
SDSS-III. The clusters are recognized for those with a richness
$R_{L\ast}\ge 12$ and a number of member galaxies candidates
$N_{200}\ge 8$ within $r_{200}$. Monte Carlo simulations show that
the false detection rate is less than 6\% for the whole sample. The
cluster detection rate is more than 95\% for clusters of
$M_{200}>1\times10^{14}~M_{\odot}$ in the redshift range of $0.05\le
z< 0.42$. We cross-match our cluster sample with previous cluster
samples. The matching rates are 50--70\% for previously known clusters
but depend on cluster richness. Clusters in our sample match more than
90\% of previously known rich clusters. Comparing with X-ray and the
Planck data, we find that the determined cluster richness is related
to the X-ray luminosity, temperature and SZ measurements.
The richer clusters have brighter BCGs. The BCGs are brighter in
higher redshift clusters. Cross-matching the BCGs with the SDSS LRGs,
we find that 25\% LRGs are BCGs of our clusters, 36\% LRGs are member
galaxies, and 63\% of the BCGs of $r_{\rm petro}<19.5$ in our
 clusters satisfy the SDSS LRG selection criteria.

\acknowledgments

We thank the referee for valuable comments that helped to improve
the paper. The authors are supported by the National Natural Science
Foundation (NNSF) of China (10833003 and 11103032).
Funding for SDSS-III has been provided by theAlfred P. Sloan
Foundation, the Participating Institutions, the National Science
Foundation, and theUSDepartment of Energy.  The SDSS-III Web site is
http://www.sdss3.org/.
SDSS-III is managed by the Astrophysical Research Consortium for the
Participating Institutions of the SDSS-III Collaboration including the
University of Arizona, the Brazilian Participation Group, Brookhaven
National Laboratory, University of Cambridge, University of Florida,
the French Participation Group, the German Participation Group, the
Instituto de Astrofisica de Canarias, the Michigan State/Notre
Dame/JINA Participation Group, Johns Hopkins University, Lawrence
Berkeley National Laboratory, Max Planck Institute for Astrophysics,
NewMexico State University, New York University, Ohio StateUniversity,
Pennsylvania State University,University of Portsmouth, Princeton
University, the Spanish Participation Group, University of Tokyo,
University of Utah,Vanderbilt University, University of Virginia,
University of Washington, and Yale University.


\end{document}